\makeatletter
\def\input@path{{tex/}}
\makeatother

\documentclass[aps,prx,superscriptaddress,reprint,longbibliography]{revtex4-2}

\usepackage{graphicx}
\usepackage{dcolumn}
\usepackage{bm}
\usepackage{natbib}
\usepackage{graphicx}
\usepackage{amsmath}
\usepackage[inline]{enumitem}
\usepackage[utf8]{inputenc}
\usepackage[T1]{fontenc}

\graphicspath{{figures/}}
\usepackage{xcolor}

\usepackage{tabularx}
\usepackage{tabulary}

\def\inn{_\mathbf{in}}
\def\out{_\mathbf{out}}

\def\rout{\mathbf{r_{out}}}
\def\rin{\mathbf{r_{in}}}
\def\xout{x_\mathrm{out}}
\def\xin{x_\mathrm{in}}
\def\uout{\mathbf{u_{out}}}
\def\uin{\mathbf{u_{in}}}

\def\tin{\theta_\mathrm{in}}
\def\ut{_{\mathbf{u}\theta}}
\def\Rxx{\mathbf{R}_{xx}}
\def\Rut{\mathbf{R}\ut}

\def\Rrr{\mathbf{R}_{\mathbf{r}\mathbf{r}}}

\def\T{\mathbf{T}}

\DeclareMathOperator\arctanh{atanh}

\newcommand*{\lauraNEW}[1]{\textcolor{black}{#1}}
\newcommand*{\laura}[1]{\textcolor{black}{#1}}
\newcommand*{\alex}[1]{\textcolor{black}{#1}}
\begin{document}

\title{Reflection matrix approach for quantitative imaging of scattering media}
\author{William Lambert}
\affiliation{Institut Langevin, ESPCI Paris, CNRS UMR 7587, PSL University, 1 rue Jussieu, 75005 Paris, France}
\affiliation{SuperSonic Imagine,
Les Jardins de la Duranne, 510 Rue Ren\'{e} Descartes, 13857 Aix-en-Provence, France}
\author{Laura A. Cobus}
\affiliation{Institut Langevin, ESPCI Paris, CNRS UMR 7587, PSL University, 1 rue Jussieu, 75005 Paris, France}
\author{Mathieu Couade}
\affiliation{SuperSonic Imagine,
Les Jardins de la Duranne, 510 Rue Ren\'{e} Descartes, 13857 Aix-en-Provence, France}
\author{Mathias Fink}
\author{Alexandre Aubry}
\email{alexandre.aubry@espci.fr}
\affiliation{Institut Langevin, ESPCI Paris, CNRS UMR 7587, PSL University, 1 rue Jussieu, 75005 Paris, France}

\date{\today}

\begin{abstract}
	We present a physically intuitive matrix approach for wave imaging and characterization in scattering media. The experimental proof-of-concept is performed with ultrasonic waves, but this approach can be applied to any field of wave physics for which multi-element technology is available. The concept is that focused beamforming enables the synthesis, in transmit and receive, of an array of virtual transducers which map the entire medium to be imaged.
	The inter-element responses of this virtual array form a focused reflection matrix from which spatial maps of various characteristics of the propagating wave can be retrieved. Here we demonstrate: (\textit{i}) a local focusing criterion that enables the image quality and the wave velocity to be evaluated everywhere inside the medium, including in random speckle, and (\textit{ii}) an highly resolved spatial mapping of the prevalence of multiple scattering, which constitutes a new and unique contrast for ultrasonic imaging. The approach is demonstrated for a controllable phantom system, and for in vivo imaging of the human abdomen. More generally, this matrix approach opens an original and powerful route for quantitative imaging in wave physics. 
\end{abstract}


\maketitle

\section{Introduction}

In wave imaging, we aim to characterize an unknown environment by actively illuminating a region and recording the reflected waves. Inhomogeneities generate back-scattered echoes that can be used to image the local reflectivity of the medium. This is the principle of, for example, ultrasound imaging~\cite{Szabo2004}, optical coherence tomography for light~\cite{Drexler2008}, radar for electromagnetic waves~\cite{Bamler1998} or reflection seismology in geophysics~\cite{Yilmaz2008}. This approach, however, rests on the assumption of a homogeneous medium between the probe and target. Large-scale fluctuations of the wave velocity in the medium can result in wavefront distortion (aberration) and a loss of resolution in the subsequent reflectivity image. Smaller-scale inhomogeneities with high concentration and/or scattering strength can induce multiple scattering events which can strongly degrade image contrast.  
In the past, numerous methods such as adaptive focusing have been implemented to correct for these fundamental issues in reflectivity imaging~\cite{Roddier1999,Booth,ODonnell1988,Mallart1994}. However, such methods are largely ineffective in situations where the focus quality inside the medium can not be determined. An extremely common example of this situation for ultrasound imaging is the presence of \textit{speckle}, the signal resulting from an incoherent sum of echoes due to randomly distributed unresolved scatterers. Speckle often dominates medical ultrasound images, making adaptive focusing difficult. 

Alternately, one can try to exploit effects which are detrimental to reflectivity imaging (such as distortion and scattering) to create different imaging modalities.  
In the ballistic regime (where single scattering dominates), the refractive index can be estimated by analyzing the distortion undergone by the wave as it passes through the medium. This is the principle of quantitative phase imaging~\cite{Park2018} in optics and computed tomography~\cite{Kak2001} in ultrasound imaging. \laura{Most such} approaches, however, require a transmission configuration, which is not practical for thick scattering media, and which is impossible for most \textit{in-vivo} or \textit{in-situ} applications in which only one side of the medium is accessible. \laura{In reflection, recent work has leveraged the relationship between the speed of sound $c$ and wavefront distortion to improve aberrated images~\cite{Ali2018,Rau2019,Chau2019} or to measure $c$~\cite{Jaeger2015SosMap}. Based on comparisons between the spatial or temporal coherence between emitted and detected signals at a transducer array, such approaches are promising for the correction of wavefront distortions 
\alex{when} single scattering dominates.} 

In the multiple scattering regime, optical diffuse tomography~\cite{Durduran2010} is a well-established technique to build a map of transport parameters. However, the spatial resolution of the resulting image is poor as it scales with imaging depth. Moreover, this approach assumes a purely multiple scattering medium, which does not exist in practice. A single scattering contribution always exists, and is furthermore typically predominant in ultrasound imaging. To evaluate the validity of such images, a local (spatially-resolved) multiple scattering rate would be a valuable observable, but is not accessible with state-of-the-art methods.

Recently, a reflection matrix approach to wave imaging was developed with the goals of: (\textit{i}) processing the huge amount of data that can now be recorded with multi-element arrays~\cite{Montaldo2009,vanPutten2010,Provost2014}, and (\textit{ii}) optimizing aberration correction~\cite{Varslot2004,Robert2008,Kang2017,Badon2019} and multiple scattering removal~\cite{Aubry2009a,Aubry2011,Kang2015,Badon2016} in post-processing. Such matrix approaches provide access to much more information than is available with conventional imaging techniques. Their recent successes suggest that access to detailed information on aberration and multiple scattering could be capitalized upon for more accurate characterization of strongly heterogeneous media. In this paper, we introduce a universal and non-invasive matrix approach for new quantitative imaging modes in reflection.

Our method is based on the projection of the reflection matrix into a focused basis \cite{Badon2016,Blondel2018}. This focused reflection matrix can be thought of as a matrix of impulse responses between virtual transducers located inside the medium~\cite{Robert2008}. These virtual transducers are created via numerical simulation of wave focusing, i.e. combining all of the backscattered echoes in such as way as to mimic focusing at a set of focal points that spans the entire medium, in both transmit and receive. While each pixel of a confocal image is associated with the same virtual transducer at emission and reception, the focused reflection matrix also contains the cross-talks between each pixel of the image, and thus holds much more information than a conventional image. Importantly, this matrix allows to probe the input-output point spread function (PSF) in the vicinity of each pixel even in speckle. A local PSF in reflection is a particularly relevant observable since it allows a local quantification of the contribution of aberration and multiple scattering to the image. More precisely, we demonstrate here the mapping of: (\textit{i}) a local focusing criterion that can then be used as a guide star for wave velocity tomography~\cite{Jaeger2015,Imbault2017,Stahli2019} in the medium, and (\textit{ii}) a spatially-resolved multiple scattering rate which paves the way towards local measurements of wave transport parameters~\cite{Aubry2008,Aubry2011,Mohanty2017} such as the absorption length  and the scattering mean free path (the mean distance between two successive scattering events). Not only are these parameters quantitative markers for biomedical diagnosis in ultrasound imaging~\cite{Suzuki1992,Sasso2010,Bamber1981,Bamber1981b,Chen1987,Duck1990} and optical microscopy~\cite{Durduran2010}, but they are also important observables for non-destructive evaluation~\cite{Schurr2011,Shajahan2014b,Zhang2016} and geophysics~\cite{Sato2012,Chaput2015,Mayor2018}. In this paper, we present the principle and first experimental proofs of concept of our approach in the context of medical ultrasound imaging. However, the concept can be extended to any field of wave physics for which multi-element technology (multiple sources/receivers which can emit/detect independently from one another) is available. \\

The paper is structured as follows: Section\ \ref{sec_ReflectionMatrixApproach} presents the concept and theoretical foundations of the focused reflection matrix. Then, experiments on a tissue-mimicking phantom are used to demonstrate proofs of concept in Section\ \ref{sec_LocalFocusingCriterion} for spatial mapping of the quality of focus and speed of sound, and in Section\ \ref{sec_MultipleScattering} for spatial mapping of multiple scattering. Perspectives for each are discussed. In Section\ \ref{sec_InvivoLiverImaging}, these techniques are applied for in vivo quantitative imaging of the human liver. 
Finally, Section\ \ref{conclusion} presents conclusions and general perspectives.

\section{Reflection Matrix approach}
\label{sec_ReflectionMatrixApproach}
\subsection{Experimental measurement}
\label{sec_subsec_ExperimentalMeasurement}

\begin{figure}[t]
\centering
\includegraphics[width=0.9\columnwidth]{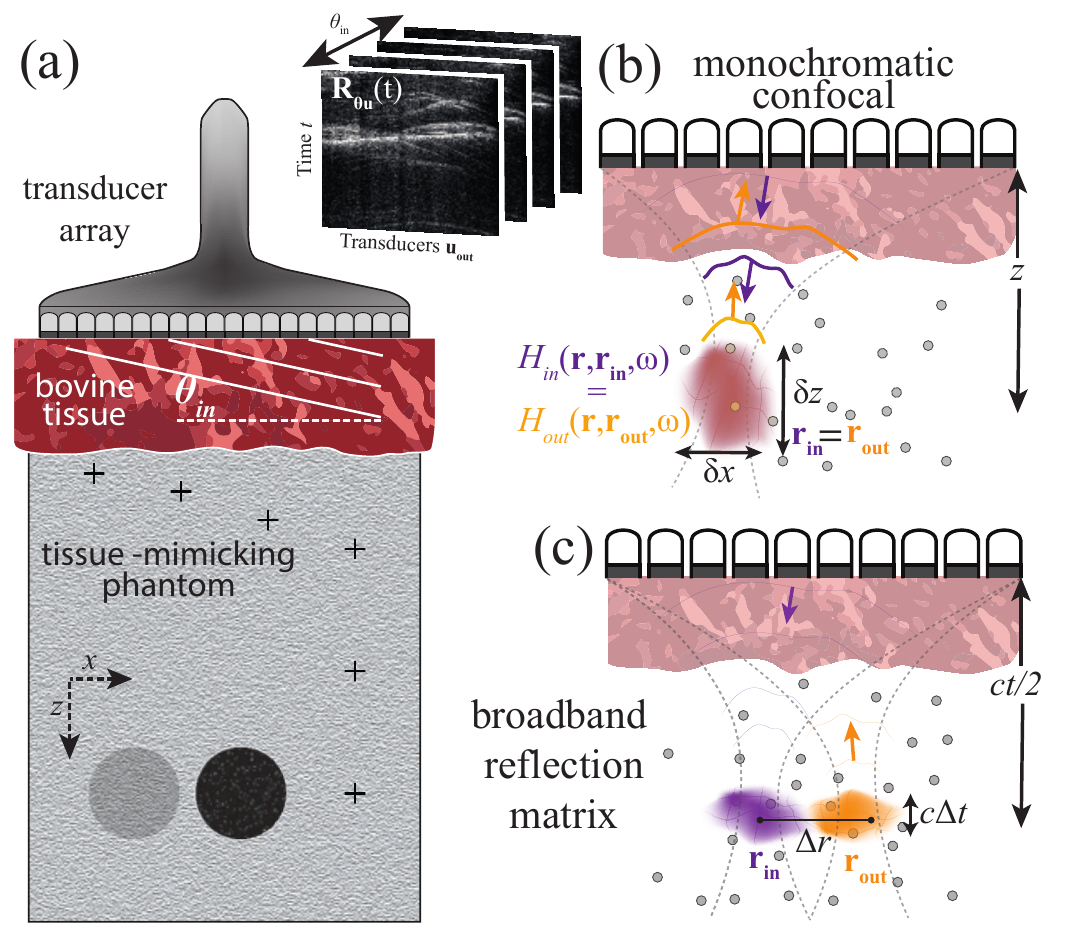}
\caption{Principle of the focused reflection matrix approach. (a) Sketch of the experimental setup for the acquisition of\ $\Rut(t)$. An ultrasonic transducer array is in direct contact with a layer of bovine tissue placed on the top of a tissue-mimicking phantom.\ $\Rut(t)$ is acquired by recording the time-dependent reflected field at each transducer element\ $\uout$, for each plane-wave illumination\ $\tin$. (b) Each pixel of a conventional image results from confocal beamforming applied to $\Rut$ in emission and reception. (c) Matrix imaging consists of performing focused beamforming to probe distinct points $\rin$ and $\rout$ in emission and reception. The set of impulse responses between such virtual transducers form a focused reflection matrix $\Rxx$ at each depth.}
\label{acq}
\end{figure}

The sample under investigation is a tissue-mimicking phantom 
\laura{composed of subresolution scatterers which generate} ultrasonic speckle characteristic of human tissue [Fig.\ \ref{acq}(a)]. \laura{The system also contains point-like specular targets are placed at regular  intervals,  and  at  larger  depths,  two  sections  of  hyperechoic cylinders, each containing a different (higher) density of unresolved scatterers.} 
A 20 mm-thick layer of bovine tissue is placed on top of the phantom and acts as both an aberrating and scattering layer. This experiment mimics the situation of \textit{in-vivo} liver imaging in which 
layers of fat and muscle tissues generate strong aberration and scattering at shallow depths. We acquire the acoustic reflection matrix experimentally using a linear ultrasonic transducer array placed in direct contact with the sample [Fig.\ \ref{acq}(a)]. The simplest acquisition sequence is to emit with one element at a time, and for each emission record with all elements the time-dependent field reflected back from the medium. This canonical basis was first used to describe the so-called time-reversal operator~\cite{Prada1994}, and is now commonly used in non destructive testing where it is referred to as the full matrix capture sequence~\cite{Holmes2005}. 
A matrix acquired in this way can be written mathematically as\ $\mathbf{R_{uu}}(t)\equiv\mathbf{R}(\uout,\uin,t)$, where $\mathbf{u}$ is the position of elements along the array, `in' denotes transmission, and `out' denotes reception. Alternately, the response matrix can be acquired using beamforming (emitting and/or receiving with all elements in concert with appropriate time delays applied to each element) to form, for example, focused beams as in the conventional B-mode~\cite{Szabo2004} or plane waves for high frame rate imaging~\cite{Montaldo2009}. To demonstrate the compatibility of our method with state-of-the-art medical technology, our data was acquired using plane-wave beamforming in emission and recording with individual elements in reception.
 
The experimental procedure is described in detail in Appendix~\ref{appA}. A set of plane waves is used to probe the medium of interest. For each plane wave emitted with an incident angle $\tin$, the time-dependent reflected wavefield is recorded by the transducers. The corresponding signals are stored in a reflection matrix\ $\Rut(t)=[ R(\uout,\tin,t)]$ [Fig.\ \ref{acq}(a)]. An ultrasound image can be formed by coherently summing the recorded echoes coming from each focal point $\mathbf{r}$, which then acts as a virtual detector inside the medium. In practice, this is done by applying appropriate time delays to the recorded signals~\cite{Montaldo2009}. The images obtained for each incident plane wave are then summed up coherently and result in a final compounded image with upgraded contrast. This last operation generates \textit{a posteriori} a synthetic focusing (i.e. a virtual source) on each focal point. The compounded image is thus equivalent to a confocal image that would be obtained by focusing waves on the same point in both the transmit and receive modes.

\subsection{Monochromatic focused reflection matrix}

We now show how all of the aforementioned imaging steps can be rewritten under a matrix formalism. The reflection matrix can actually be defined in general as containing responses between one or two mathematical bases. The bases implicated in this work are: (\textit{i}) the recording basis which here corresponds to the transducer array, (\textit{ii}) the illumination basis which is composed of the incident plane waves, and (\textit{iii}) the focused basis in which the ultrasound image is built. In the frequency domain, simple matrix products allow ultrasonic data to be easily projected from the illumination and recording bases to the focused basis where local information on the medium proprieties can be extracted.

Consequently, a temporal Fourier transform should be first applied to the experimentally acquired reflection matrix to obtain\ $\Rut(\omega)$, where\ $\omega=2\pi f$ is the angular frequency of the waves.	The matrix $\mathbf{R_{u\theta}}(\omega)$ can be expressed as follows:
\begin{equation}
\label{Rut}
\mathbf{R_{u\theta}}(\omega)=\mathbf{G}^\top(\omega) \times \mathbf{\Gamma} \times  \mathbf{T} ,
\end{equation}
where the matrix $\mathbf{\Gamma}$, defined in the focused basis, describes the scattering process inside the medium. In the single scattering regime, $\mathbf{\Gamma}$ is diagonal and its elements correspond to the medium reflectivity $\gamma(\mathbf{r})$. $\mathbf{T}=[T(\mathbf{r},\theta)]$ is the transmission matrix between the plane wave and focused bases. Each column of this matrix describes the incident wavefield induced inside the sample by a plane wave of angle $\theta$. $\mathbf{G}=[G(\mathbf{u},\mathbf{r})]$ is the Green's matrix between the transducer and focused bases. Each line of this matrix corresponds to the wavefront that would be recorded by the array of transducers along vector $\mathbf{u}$ if a point source was introduced at a point $\mathbf{r}=(x,z)$ inside the sample.

The holy grail for imaging is to have access to these transmission and Green's matrices. Their inversion, pseudo-inversion, or more simply their phase conjugation can enable the reconstruction of a reliable image of the scattering medium, thereby overcoming the aberration and multiple scattering effects induced by the medium itself. However, direct measurement of the transmission and Green's matrices $\mathbf{T}$ and $\mathbf{G}$ would require the introduction of sensors inside the medium, and therefore these matrices are not accessible in most imaging configurations.
Instead, sound propagation from the plane-wave or transducer bases to the focal points is usually modelled assuming a homogeneous speed of sound $c$. In this case, the elements of the corresponding free-space transmission matrix  $\mathbf{T_{0}}(\omega)$ are given by
\begin{equation}
\label{Gzp_eq}
T_0\left(\theta,\mathbf{r},\omega\right)=\exp{\left[i k \left(z \cos\theta + x\ \sin \theta \right)\right]} ,
\end{equation}
where\ $x$ and\ $z$ describe the coordinates of $\mathbf{r}$ in the lateral and axial directions, respectively [Fig.\ \ref{acq}(a)], and $k=\omega/c$ is the wave number. The elements of the free-space Green's matrix\ $\mathbf{G_0}(\omega)$ are the 2D Green's functions between the transducers and the focal points~\cite{Watanabe}
\begin{equation}
\label{Gz_eq}
G_0\left(\rout,\uout,\omega\right)  =- \frac{i}{4}  \mathcal{H}_0^{(1)}\left (k|\rout-\uout| \right ) ,
\end{equation}
where $\mathcal{H}_0^{(1)} $ is the Hankel function of the first kind. $\mathbf{T_{0}}(\omega)$ and 	$\mathbf{G_{0}}(\omega)$ can be used to project the reflection matrix $\mathbf{R_{u\theta}}(\omega)$ into the focused basis. Based on Kirchhoff's diffraction theory~\cite{Goodman1996}, 
such a double focusing operation can be written as the following matrix product
\begin{equation}
\Rrr (\omega)
=  \mathbf{{G_0^*}}\left(\omega\right)\times\Rut(\omega)\times \mathbf{T_0^{\dagger}}\left(\omega\right),
\label{projRrr}
\end{equation}
where the symbol $*$ and $\dagger$ stand for phase conjugate and transpose conjugate, respectively. The matrices $\mathbf{T_0^{\dag}}$ and $\mathbf{G_0^*}$ contain the phase-conjugated wavefronts that should be applied at emission and reception in order to project the reflection matrix into the focused basis both at input ($\rin$) and output ($\rout$). Equation (\ref{projRrr}) thus mimics focused beamforming in post-processing in both emission and reception. 
Each coefficient of $\mathbf{R_{rr}}=[R(\mathbf{r_\text{out}}, \rin)]$ is the impulse response between a virtual source at point $\rin$ and a virtual detector at $\rout$ [Fig.\ \ref{acq}(c)]. 

The aberration issue in imaging can be investigated by expressing the matrix $\mathbf{R}_{\mathbf{r}\mathbf{r}}$ mathematically using Eqs.~(\ref{Rut}) and (\ref{projRrr}): 
\begin{equation}
\label{RrrMatrix}
\mathbf{R}_{\mathbf{r}\mathbf{r}}(\omega)  = \mathbf{H}^{\top}\out (\omega)  \times \mathbf{\Gamma} \times \mathbf{H}\inn (\omega),
\end{equation}
where
\begin{equation} 
\label{H}
\mathbf{H}\inn (\omega)=\mathbf{T}(\omega)\times \mathbf{T_0^{\dagger}}(\omega) 
\end{equation}
and 	
\begin{equation} 
\label{Hout}
\mathbf{H}\out(\omega)=\mathbf{G}(\omega)\times \mathbf{G_0^{\dagger}}(\omega) 
\end{equation}	
are the input and output focusing matrices, respectively [Fig.\ \ref{acq}(b)]. Each column of $\mathbf{H}\inn=[H\inn(\mathbf{r},\rin)]$ and $\mathbf{H}\out=[H\out(\mathbf{r},\rout)]$ corresponds to the transmit and receive PSFs, i.e. the spatial amplitude distribution of the input and output focal spots. Their support defines the characteristic size of each virtual source at $\rin$ and detector at $\rout$ [Fig.\ \ref{acq}(b)]. In the absence of aberration, the transverse and axial dimension of these focal spots, $\delta x_0$ and $\delta z_0$, are only limited by diffraction~\cite{Born}:
\begin{equation}
\delta  x_0=\frac{\lambda}{2 \sin \beta}  \, \text{,    } \delta z_0 =   \frac{ 2\lambda }{ \sin^2 \beta},
\end{equation}
where $\beta$ is the maximum angle of wave illumination or collection by the array and $\lambda$ the wavelength. In the presence of aberration, i.e. if the velocity model is inaccurate, there is a mismatch between the transmission and Green's matrices, $\mathbf{T}(\omega)$ and $\mathbf{G}(\omega)$, and their free-space counterparts, $\mathbf{T_0}(\omega)$ and $\mathbf{G_0}(\omega)$. The focusing matrices, $\mathbf{H}\inn$ and $\mathbf{H}\out$, are far from being diagonal [Eqs.(\ref{H})-(\ref{Hout})]. The corresponding PSFs are strongly degraded and the virtual transducers can overlap significantly. A better model of wave propagation is thus needed to overcome aberrations and restore diffraction-limited PSFs.  In Sec.~\ref{tomo}, we will show how a multilayer
model can be used to reach a better estimate of $\mathbf{T}$ and $\mathbf{G}$ in the experimental configuration depicted in Fig.~\ref{acq}(a). 

\subsection{Broadband focused reflection matrix}

\begin{figure*}[t]
\centering
\includegraphics[width=\textwidth]{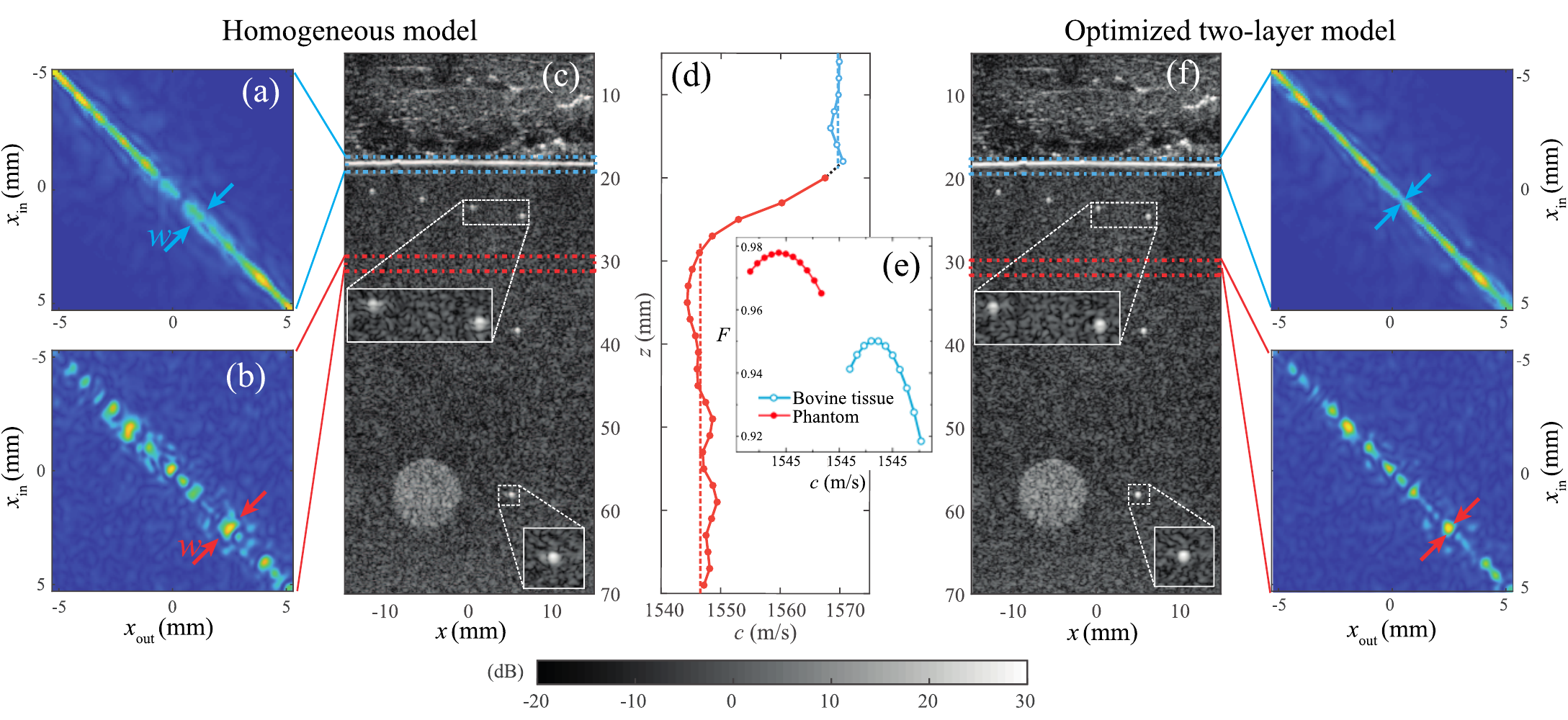} 
\caption{Matrix approach applied to wave velocity mapping of the bovine tissue/phantom system described in Fig.~\ref{acq}(a). (a,b) The matrix $\Rxx$ is displayed at depth $z=18$ mm and $30$ mm, respectively, assuming an homogeneous wave velocity model ($c=1542$ m.s$^{-1}$). The local image resolution $w$ is extracted from each antidiagonal of $\Rxx$.
	(c) Corresponding ultrasound image built from the confocal elements of $\Rxx$.  
	(d) The optimized wave velocities are displayed versus depth for the bovine tissue, found using a homogeneous model (blue open symbols), and the phantom, found using a bi-layer model (red solid symbols). (e) The focusing criterion $F$, averaged over the depth ranges [17.4, 19.4]~mm (blue circles) and [30,32]~mm (red disks), is displayed versus the wave velocity hypothesis\ $c$. (f) Ultrasound image built from the confocal elements of $\Rxx$ using the optimized wave velocity model ($c_t=1573$~m/s,\ $c_p=1546$~m/s). (g,h) Corresponding reflection matrices $\Rxx$ are shown for depths $z=18$~mm and $z=30$~mm. }
\label{qualityoffocus_img}
\end{figure*}
For broadband imaging, we can restrict our study to pairs of virtual transducers,\ $\rin=(x_{\text{in}},z)$ and $\rout=(x_{\text{out}},z)$, located at the same depth\ $z$. Furthermore, an inverse Fourier transform of the corresponding sub-matrices $\Rxx(z,\omega)=[R(x_\text{out},x_\text{in},z,\omega)]$ should be performed in order to recover the excellent axial resolution of ultrasound images. For direct imaging, only echoes at the ballistic time ($t=0$ in the focused basis) are of interest. This ballistic time gating can be performed via a coherent sum of $\Rxx(z,\omega)$ over the frequency bandwidth $\delta \omega$. A broadband focused reflection matrix\ $\Rrr(z)$ is thus obtained at each depth $z$:
\begin{equation}
\label{Rrr_focused_eq}
\Rxx (z) = \int^{\omega_{+}}_{\omega_{-}} d\omega \Rxx (z,\omega) ,
\end{equation}
where\ $\omega_{\pm}=\omega_c \pm \delta \omega/2$ and\ $\omega_c$ is the central frequency. Each element of\ $\Rxx(z)$ contains the signal that would be recorded by a virtual transducer located at\ $\rout=(x_\text{out},z)$ just after a virtual source at\ $\rin=(x_\text{in},z)$ emits a pulse of length $\Delta t= \delta \omega^{-1}$ at the central frequency $\omega_c$. The broadband focusing operation of Eq.\ (\ref{Rrr_focused_eq}) gives virtual transducers which now have a greatly reduced axial dimension $\delta z \sim c \Delta t$ [Fig.\ \ref{acq}(c)].  

Figures \ref{qualityoffocus_img}(a) and (b) display $\Rxx$ at the bovine tissue/phantom interface ($z=18$ mm) and in the phantom ($z=30$ mm), respectively. In both cases, most of the signal is concentrated around the diagonal. This indicates that single scattering dominates at these depths~\cite{Badon2016}, since a singly-scattered wavefield can only originate from the point which was illuminated by the incident focal spot. In fact, the elements of\ $\Rrr$ which obey $\rin=\rout$ hold the information which would be obtained via multifocus (or confocal) imaging, in which transmit and receive focusing are performed at the same location for each point in the medium.
A line of the ultrasound image can thus be directly deduced from the diagonal elements of $\Rxx(z)$, computed at each depth: 
\begin{equation}
\label{imcalc}
\mathcal{I}\left(\mathbf{r}\right)\equiv \left|R\left(x,x,z\right)\right|^2.
\end{equation}
The corresponding image is displayed in Fig.\ \ref{qualityoffocus_img}(c). It is equivalent to the coherent compounding image computed via delay-and-sum beamforming of the same data set~\cite{Montaldo2009}, constituting a validation of our matrix approach for imaging.

Interestingly, the matrix $\Rrr$ contains much more information than a single ultrasound image. In particular, focusing quality can be assessed by means of the off-diagonal elements of $\Rxx$. To understand why this is, $\Rxx$ shall be expressed theoretically. To that aim, a time-gated version of Eq.~(\ref{RrrMatrix}) can be derived:
\begin{equation}
\label{RxxMatrix}
\Rxx(z)=\mathbf{H^{\top}_\text{out}}(z)\times \mathbf{\Gamma}(z) \times \mathbf{H_\text{in}}(z),
\end{equation}
where $\mathbf{H\inn}(z)$, $\mathbf{\Gamma}(z)$ and $\mathbf{H}\out(z)$ are the time-gated sub-matrices of $\mathbf{H}\inn$ [Eq.~(\ref{H})], $\mathbf{\Gamma}$ 
[Eq.~(\ref{Rut})] and $\mathbf{H}\out$ [Eq.~(\ref{Hout})] at depth $z=ct/2$ and central frequency $\omega_c$.
In the single scattering regime and for spatially- and frequency-invariant aberration, the previous equation can be rewritten in terms of matrix coefficients as follows:
\begin{eqnarray}
\label{Rrr_intermsofh_eq}
R({x_\text{out}},{x_\text{in}},z)=\int dx & &  H_\text{out}(x-x_\text{out},z) \nonumber \\
& &\times  \gamma(x,z) H_\text{in}(x-x_\text{in},z) .
\end{eqnarray}
This last equation confirms that the diagonal coefficients of $\Rxx(z)$, i.e. a line of the ultrasound image, result from a convolution between the sample reflectivity $\gamma$ and the confocal PSF $H_\text{in} \times H_\text{out}$. As we will see, access to the off-diagonal elements of $\Rxx$ will allow our analysis of the experimental data to go far beyond a simple image of the reflectivity. In particular, off-diagonal elements can be used to extract the input-output PSF in the vicinity of each focal point, which will lead to a local quantification of the focusing quality.

\section{Local focusing criterion}
\label{sec_LocalFocusingCriterion}

In this section, we detail how an investigation of the off-diagonal 	points in $\Rxx$ can directly provide a focusing quality criterion for any pixel of the ultrasound image. To that aim, the relevant observable is the mean intensity profile along each antidiagonal of $\Rxx$: 
\begin{equation}
\label{IrDr_fromRrDr}
I(\mathbf{r},\Delta x )=\left \langle \left | R(x +\Delta x /2,x-\Delta x /2,z) \right |^2 \right \rangle ,
\end{equation}
where $\langle\cdots\rangle$ denotes an average over the pairs of points $\rin=(\xin,z)$ and $\rout=(\xout,z)$ which share the same midpoint $\mathbf{r}=(\rout+\rin)/2$, and $\Delta x=(x\out-x\inn)$ is the relative position between those two points.	
We term $I(\mathbf{r},\Delta x )$ the \textit{common-midpoint intensity profile}.  Whereas $\mathcal{I}\left(\mathbf{r}\right)$ [Eq.\ (\ref{imcalc})] only contains the confocal intensity response from an impulse at point $\mathbf{r}$,  $I(\mathbf{r},\Delta x )$ is a measure of the spatially-dependent intensity response to an impulse at $\mathbf{r}$. 
This means that whatever the scattering properties of the sample,  $I(\mathbf{r},\Delta x )$ allows an estimation of the input-output PSFs. However, its theoretical expression differs slightly depending on the characteristic length scale $l_{\gamma}$ of the reflectivity $\gamma(\mathbf{r})$ at the ballistic depth and the typical width $\delta x$ of the input and output focal spots.

In the specular scattering regime [$l_{\gamma}>>\delta x$, see Fig.\ \ref{qualityoffocus_img}(a)],  the common-midpoint intensity is directly proportional to the convolution between the coherent input and output PSFs, $H_\text{in}$ and $H_\text{out} $ (see Appendix \ref{appB}):
\begin{equation}
\label{IrDr_spec}
I(\mathbf{r},\Delta x )=   | \gamma(\mathbf{r})|^2  \times | \left (  H_\text{in}  \ast H_\text{out} \right) (\Delta \mathbf{r}) |^2,
\end{equation}
where the symbol $\ast$ stands for convolution. However, in ultrasound imaging, scattering is more often due to a random distribution of unresolved scatterers. In this speckle regime [$l_{\gamma}<\delta x$, see Fig.\ \ref{qualityoffocus_img}(b)], the common midpoint intensity $I(\mathbf{r},\Delta x )$ is directly proportional to the convolution between the incoherent input and output PSFs, $|H_\text{in}|^2$ and $|H_\text{out}|^2$ (see Appendix \ref{appB}):
\begin{equation}
\label{IrDr_fromRrDr2}
I(\mathbf{r},\Delta x )=  \left \langle | \gamma(\mathbf{r})|^2 \right \rangle \times  \left ( |H_\text{in}|^2  \ast |H_\text{out}|^2 \right) (\Delta \mathbf{r}).
\end{equation}
The ensemble average in Eq.\ \ref{IrDr_fromRrDr2} implies access to several realizations of disorder for each image, which is often not possible for most applications. In the absence of multiple realizations, a spatial average over a few resolution cells is required to smooth intensity fluctuations due to the random reflectivity of the sample while keeping a satisfactory spatial resolution. To do so, a spatially averaged intensity profile $ I_{\text{av}} (\mathbf{r},\Delta x)$ is computed at each point $\mathbf{r}$ of the field of view, such that
\begin{equation}
\label{avIrDr_fromRrDr}
I_{\text{av}} (\mathbf{r},\Delta x)=\left \langle I(\mathbf{r'},\Delta x) \right \rangle_{(\mathbf{r'}-\mathbf{r}) \in \mathcal{A}}
\end{equation}
where the symbol $\langle \cdots \rangle $ denotes an average over the set of focusing points $\mathbf{r'}$ contained in an area $\mathcal{A}$ centered on $\mathbf{r}$. The compromise between intensity fluctuations and spatial resolution guided our choice of a 7.5 mm-diameter disk for $\mathcal{A}$.

Whatever the scattering regime, the averaged common-midpoint intensity profile $I_{\text{av}} (\mathbf{r},\Delta x)$ is a direct indicator of the focusing resolution at each point\ $\mathbf{r}$ of the medium. One can then build a local focusing parameter 
\begin{equation}
\label{F}
F(\mathbf{r})=w_0(\mathbf{r})/w(\mathbf{r}),   
\end{equation}
where\ the input-output focusing resolution $w(\mathbf{r})$ is defined as the full width at half maximum (FWHM) of $I_{\text{av}}(\mathbf{r},\Delta x )$, and $w_0(\mathbf{r})$ is a reference value based on the theoretical diffraction limit for a homogeneous medium. This parameter is bounded between 0 ($w>>w_0$, bad focusing) and 1 ($w=w_0$, perfect focusing). $F(\mathbf{r})$ is the equivalent in the focused basis of the coherence or focusing factor originally introduced by Mallart and Fink~\cite{Mallart1994} in the transducer basis. The definition of a focusing parameter in the focused basis offers an important advantage in that the wave focusing quality and the image resolution can now be probed locally.
\begin{figure}[b]
\centering
\includegraphics[width=\columnwidth]{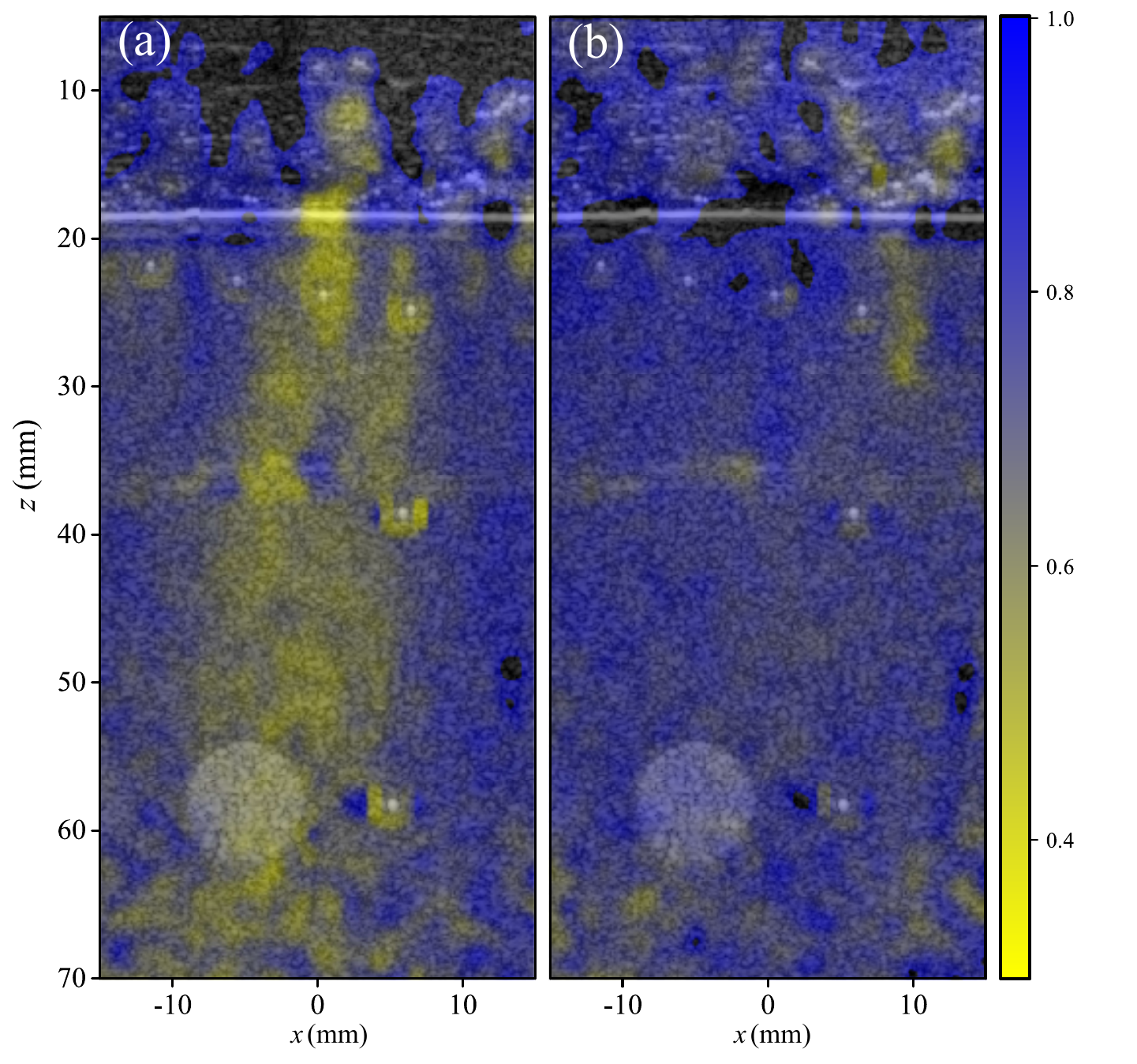}
\caption{Maps of local focusing parameter $F(\mathbf{r})$ for the bovine tissue and phantom system, superimposed over the echographic image of Fig.\ \ref{qualityoffocus_img}(c). (a) The homogeneous model with a constant speed of sound ($c=1542$~m/s) results in a poor quality of focus in some areas. (b) The two-layer model used to construct $\Rrr$ results in close to ideal focus quality throughout the image.}	
\label{ResMap_fig}
\end{figure}	

Figure\ \ref{ResMap_fig}(a) shows the focusing criterion $F(\mathbf{r})$ calculated for the bovine tissue/phantom system [corresponding to the reflectivity image of Fig.\ \ref{qualityoffocus_img}(c)]. The reference resolution $w_0(\mathbf{r})$ has been computed under a speckle scattering hypothesis [Eq.~(\ref{random})]. The poor quality of focus over a large part of the image can be attributed to the fact that the presence of the bovine tissue layer was not taken into account in our (homogeneous) model of the system [Eqs.\ (\ref{Gzp_eq}) --\ (\ref{projRrr})]. Fortunately, as discussed in the following section, the focused reflection matrix approach enables the determination of a more accurate model for this \textit{a priori} unknown medium.

\subsection{\label{tomo}Wave velocity mapping}
\label{sec_WaveVelocityMapping}

The ability to locally probe the focus quality offers enormous advantages for local characterization of heterogeneous media, in particular for a quantitative mapping of their refractive index, or more specifically, as shown in the following, the speed of sound.

We begin by observing the focusing criterion $F(\mathbf{r})$ at the bovine tissue-phantom interface as a function of the wave velocity $c_t$ assumed in the bovine tissue. The result is displayed in Fig.\ \ref{qualityoffocus_img}(e). The corresponding focusing criterion is optimized for\ $c_t=1573 $~m/s. Fig.\ \ref{qualityoffocus_img}(g) shows the reflection matrix $\Rxx$ obtained by considering this optimized wave velocity. The comparison with the original matrix displayed in Fig.\ \ref{qualityoffocus_img}(a) shows a narrowing of the input-output PSFs along the antidiagonal of\ $\Rxx$; the focusing resolution\ $w(\mathbf{r})$ is now much closer to the diffraction limit\ $w_0$ due to the use of the optimized\ $c_t$.

This approach works for a reasonably homogeneous medium (the tissue layer). However, our assumption of a homogeneous wave velocity model does not conform to the bi-layer system under experimental investigation [Fig.\ \ref{acq}(a)]. To probe more deeply into the system, we extend our approach to model a multilayer medium. Using the ultrasound image [Fig.\ \ref{qualityoffocus_img}(c)] as an approximate guide, we define two layers: one at $z=0-18$~mm with our measured $ c_t $, and a second for depths below $z=18$~mm with unknown wave velocity $c_p$. New transmission and Green's  matrices, $\mathbf{{T_1}}$ and $\mathbf{{G_1}}$, are computed using this two-layer wave velocity model. A new reflection matrix $\mathbf{R^\prime_{xx}}$ is then built via
\begin{equation}
\mathbf{R^\prime_{xx}}(z) = \int_{\omega_{-} }^{\omega_{+}} d\omega \mathbf{{G_1^*}}\left(z,\omega\right)\times\Rut(\omega)\times \mathbf{T_1^{\dagger}}\left(z,\omega\right).
\end{equation}
Figure\ \ref{qualityoffocus_img}(h) displays $\mathbf{R^\prime_{xx}}$ at depth $z=30$~mm. The corresponding focusing criterion $F(\mathbf{r})$, averaged over the full width of the image and a $2$~mm range of depths, is shown as a function of the phantom speed of sound hypothesis\ $c_p$ in Fig.\ \ref{qualityoffocus_img}(e). The optimization of $F$ yields a quantitative measurement of the speed of sound in the phantom: $c_p=1546$~m/s. 

To build an entire profile of wave velocity throughout the medium, the $F-$optimization is repeated for each depth. The resulting depth-dependent velocity estimate is shown in Fig.\ \ref{qualityoffocus_img}(d). The presence of two layers can be clearly seen, corresponding to the bovine tissue with mean wave velocity\ $\langle c_t \rangle = 1570$~m/s ($z<18 $~mm), and the phantom with $\langle c_p \rangle=1547$~m/s ($z>30 $~mm). These values are in excellent agreement with the manufacturer's specification for the phantom ($c_p=1542\pm 10$~m/s) and the speed of sound estimated from the travel time of the pulse reflected off of the tissue/phantom interface ($ c_t \approx\ 1573$~m/s). At depths just below the interface between the two layers, the measurement of $c_p$ appears to be less precise. This effect can be explained by the fact that the measurement error $\Delta c_p/c_p$ on the wave velocity scales as the inverse of $z_p$, the depth of the focal plane from the phantom surface [see Appendix \ref{appD}, Eq.~\ref{errorF_appendix}]:
\begin{equation}
   \left ( \frac{\Delta c_p}{c_p} \right )^2 \sim   \frac{1}{(k_p z_p)^2  }\frac{\sin \beta }{   \arctanh (\sin \beta) -  \beta^2/\sin \beta}\frac{\Delta F}{F},	    \end{equation}
with $k_p=\omega_c/c_p$. As the precision with which the focusing criterion $F$ can be measured is $\Delta F/F \sim 5\times 10^{-4}$ [see Fig.~\ref{qualityoffocus_img}(e)], a precision of $\Delta c_p \sim 5$ m/s for the wave velocity in the second layer (the phantom) will only be reached for $z_p\gtrsim 10$~mm. This value is in qualitative agreement with the axial resolution of the wave velocity profile displayed in Fig.~\ref{qualityoffocus_img}(d).

Figure\ \ref{qualityoffocus_img}(f) shows the ultrasound image deduced from the confocal elements of $\mathbf{R^\prime_{xx}}$. Compared with the homogeneous model [Fig.\ \ref{qualityoffocus_img}(c)], it can be seen by eye that the two-layer model slightly improves the imaging of bright targets, but that there is no clear difference in areas of speckle. However, the result for $F(\mathbf{r})$ after optimization with the two-layer model shows that a significant improvement in quality of focus has been obtained with the two-layer model (Fig.\ \ref{ResMap_fig}). The significance of this result is that, in regions of speckle, $F(\mathbf{r})$ is far more sensitive than image brightness to the quality of focus and speed of sound. As most state of the art methods for speed of sound measurement are based on image brightness~\cite{Mehta2008,Dasarathy2009,Zubajlo2018}, $F(\mathbf{r})$ thus constitutes an important new metric for speed of sound measurement in heterogeneous media.

\subsection{Discussion}

The study of the focused reflection matrix yields a quantitative, local focusing criterion. To our knowledge, this is the first demonstration of such spatial mapping of focus quality (e.g. Fig.\ \ref{ResMap_fig}). While this information is useful to evaluate the reliability of the associated ultrasound images, \laura{it has even more potentional for therapeutic ultrasound methods which rely on precisely focused beams for energy delivery, such as high-intensity focused ultrasound (HIFU)~\cite{Kennedy2005}, ultrasound neuromodulation~\cite{Blackmore2019}, and histotripsy~\cite{Macoskey2018}}.

Going further, the local focusing parameter\ $F(\mathbf{r})$ constitutes a sensitive guide star to map the wave velocity of an inhomogeneous scattering medium. The perspective of this work will be to go beyond a depth profile of the wave velocity to map its variations in 2D (1D probe) or 3D (2D array). In this respect, we shall mention the recent work of Jaeger \textit{et al.}~\cite{Jaeger2015,Stahli2019} that investigated the local phase change at a point when changing the transmit beam steering angle. Looking at this local phase change under a matrix formalism would be a way to make the best of the two approaches. An inverse problem would then have to be solved to retrieve a map of the local phase velocity~\cite{Jaeger2015,Stahli2019}. Again, a matrix formalism could be relevant to optimize this inversion.
	
In the same context, we would also like to mention the work of Imbault \textit{et al.}~\cite{Imbault2017} that investigated the correlations of the reflected wavefield in the transducer basis for a set of focused illuminations. Combined with a time reversal process consisting in iteratively synthesizing  a virtual reflector in speckle~\cite{Montaldo2011}, the wave speed was measured by maximizing a focusing criterion based on the spatial correlations of the reflected wavefield~\cite{Mallart1994}. The downside of this approach is that the construction of the virtual reflector requires that the focusing algorithm be iterated several times before the guide-star becomes point-like. Moreover, the whole process should be both averaged and repeated over each isoplanatic patch of the image~\cite{Dahl2005}, which limits the spatial resolution and the practicability of such a measurement.  

Inspired by previous works~\cite{Mallart1994,Robert2008,Montaldo2011,Imbault2017}, novel potential applications can be imagined for $F(\mathbf{r})$. It could, for instance, be used as a guide star for a matrix correction of aberration. Based on its maximization, the goal would be to converge towards the best estimators of the transmission matrices\ $\mathbf{T}$ and\ $\mathbf{G}$. An inversion or pseudo-inversion of these matrices would then lead to an optimized image whose resolution would be only limited by diffraction. \\

The developments presented thus far have been based on a single scattering assumption. However, multiple scattering is \laura{often} far from being negligible in real-life ultrasound imaging, whether it be for example in soft human tissues~\cite{Aubry2011} or coarse-grain materials~\cite{Shajahan2014b}. In the following, we show how the reflection matrix approach suggests a solution for the multiple scattering problem, and how it can furthermore be exploited to create a new contrast for ultrasound imaging. 

\section{Multiple scattering}
\label{sec_MultipleScattering}
	
\begin{figure}
\centering
\includegraphics[width=\columnwidth]{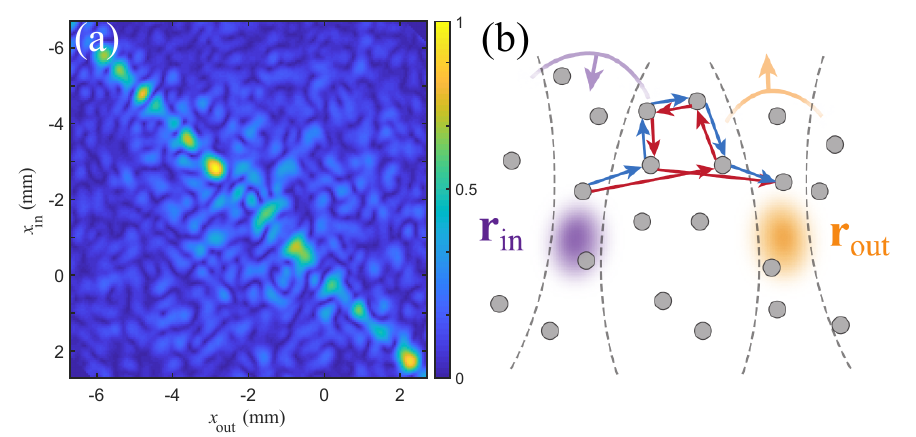}	
\caption{(a) The modulus of matrix $\Rxx$ is shown at depth $z=44.5$ mm. Signals from multiple scattering can be seen at elements far from the diagonal. (b) Sketch of multiple scattering paths (red or blue path) involved in the matrix $\Rxx$. The constructive interference between reciprocal paths (red and blue paths) occurs only when \ $|\mathbf{r_\text{out}} - \mathbf{r_\text{in}}|<\delta x$ (CBS).}
\label{MS_sketch_Rrr_fig}
\end{figure}

In the previous sections, we developed a local focusing criterion by considering the near-confocal elements of\ $\Rxx$. We now turn our attention to the points which are farther from the confocal elements, i.e. for which\ $|\xout-\xin|> w$. Figure\ \ref{MS_sketch_Rrr_fig}(a) shows\ $\Rxx$ at a depth of\ $z=60$~mm. Signal can clearly be seen at points far from the diagonal. Because each matrix\ $\Rxx$ is investigated at the ballistic time ($t=2z/c$), the only possible physical origin of echoes between distant virtual transducers is the existence of multiple scattering paths occurring at depths shallower than the focal depth, as sketched in Fig.\ \ref{MS_sketch_Rrr_fig}(b). In this section, we will see that a significant amount of multiple scattering takes place in our bovine tissue/phantom system. Signal from such multiple scattering processes has traditionally been seen as a nightmare for classical wave imaging, as it presents as an incoherent background which can greatly degrade image contrast. However, because they are extremely sensitive to the micro-architecture of the medium, multiply scattered waves can be a valuable tool for the characterization of scattering media~\cite{Aubry2008,Aubry2011,Mohanty2017}. In the following, we show how our matrix approach enables the measurement of a local multiple scattering rate for each pixel of the ultrasound image. This multiple scattering rate can be directly related to the concentration of scatterers, paving the way towards a novel contrast for ultrasound imaging.

\subsection{Multiple scattering in the focused basis}

For our experimental configuration, multiple scattering can be investigated by examining the averaged spatial intensity profiles\ $I_{\text{av}}(\mathbf{r},\Delta x)$ [Eq.\ (\ref{avIrDr_fromRrDr})]. Each intensity profile is composed of three contributions: \\

\begin{enumerate}
\item The single scattering component, $I_S$. Signals from single scattering mainly lie along the near-confocal elements of $\Rxx$ [$\Delta r < w(\mathbf{r})$]. This is the contribution that has been investigated in the previous sections. 
\item The multiple scattering component, $I_M$. This contribution can be split into two terms: An \textit{incoherent} part which corresponds to interferences between waves taking different paths through the medium, and a \textit{coherent} part which corresponds to the interference of waves with their reciprocal counterparts [see the blue and red paths in Fig.\ \ref{MS_sketch_Rrr_fig}(b). Referred to as coherent backscattering (CBS), this interference phenomenon results in an enhancement (of around two) in intensity at exact backscattering. Originally discovered in the plane wave basis~\cite{Kuga1984,VanAlbada1985b,Wolf1985,Akkermans1988c}, this phenomenon also occurs in a point-to-point basis, whether the points be real sensors ~\cite{Bayer1993a,Tourin1997,Larose2004} or created via focused beamforming~\cite{Aubry2007b,Aubry2008}. In the point-to-point basis, contributions from multiple scattering give to the backscattered intensity profile the following shape: a narrow, steep peak (the CBS peak) in the vicinity of the source location [$\Delta x < w(\mathbf{r})$], which sits on top of a wider pedestal (the incoherent contribution). 
\item Electronic noise, $I_N$. These contribution can decrease the contrast of an ultrasound image in the same way as $I_M$. Noise contributes to a roughly constant background level to the backscattered intensity profiles $I_{\text{av}}(\mathbf{r},\Delta x)$.
\end{enumerate}

\begin{figure*}[t]
\centering
\includegraphics[width=\textwidth]{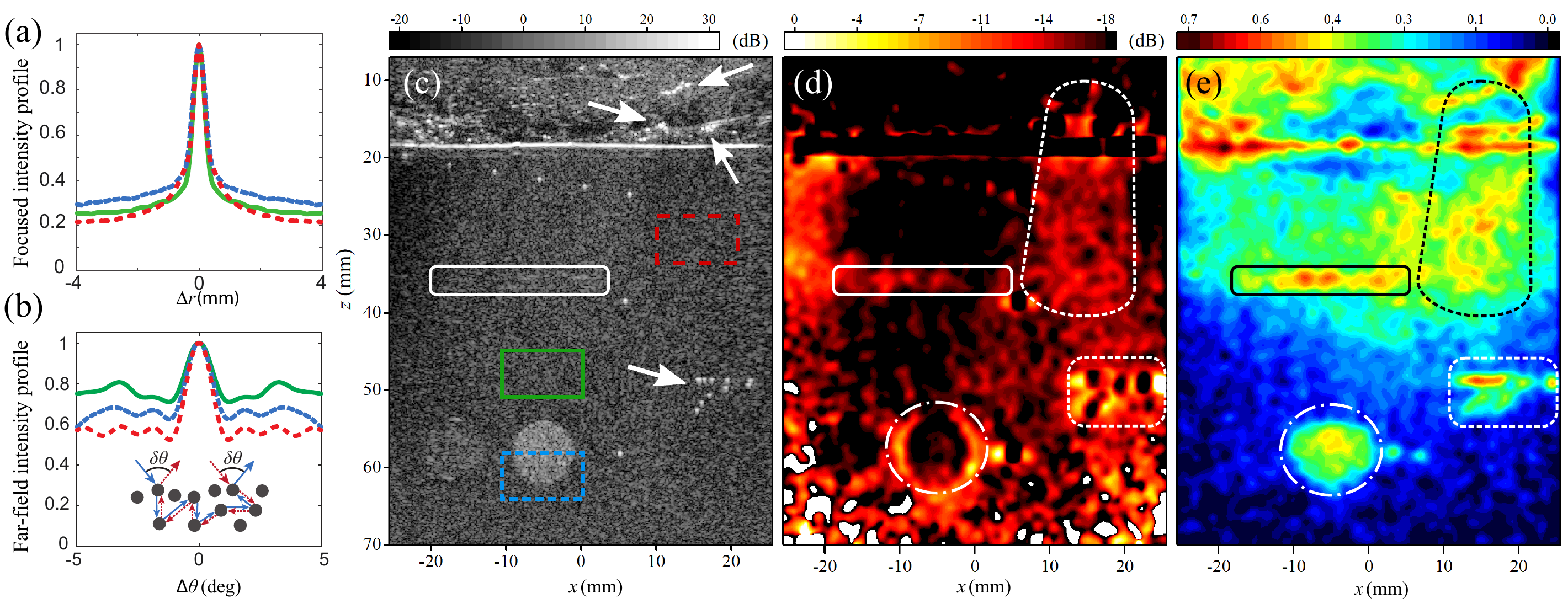}	
\caption{Local mapping of multiple scattering.
 Normalized mean intensity profiles are displayed for (a) the focused basis $I_{\text{av}}(\mathbf{r},\Delta x)$ and (b) the far-field $I_{\text{av}}(\mathbf{r},\Delta \theta)$,  are displayed for the different areas highlighted in (c) the corresponding ultrasound image (formed using the two-layer model of Section\ \ref{tomo}). Maps of multiple scattering rates (d) $\rho(\mathbf{r})$ [Eq.~\ref{gamma_defeq}] and (e) $\epsilon(\mathbf{r})$ [Eq.~\ref{epsilon_defeq}] are shown, superimposed on the ultrasound image. Some structures in the bovine tissue layer (indicated by white arrows) cause a significant amount of multiple scattering to occur behind them (outlined by dashed lines). The solid lines outline the image area that suffers from artifacts due to the double reflection event between the probe and the bovine tissue-phantom interface.}	
\label{MSfig}
\end{figure*}	
To estimate the level of each contribution, the relevant observables are the mean confocal intensity ($I_{\text{on}}$) and off-diagonal intensity ($I_{\text{off}}$) of $\Rxx$.
The confocal intensity $I_{\text{on}}$ is given by
\begin{equation}
\label{Ion}
I_{\text{on}}(\mathbf{r})= I_{\text{av}}( \mathbf{r},\Delta x=0)  = I_S(\mathbf{r}) + 2I_M(\mathbf{r})+ I_N ,
\end{equation}	
where the factor of $2$ accounts for the CBS enhancement of the multiple scattering intensity at the source location.   
$I_{\text{off}}$ is the sum of the multiple scattering incoherent background and of the additive noise component:
\begin{equation}
\label{Ioff}
I_{\text{off}}(\mathbf{r})= \langle I_{\text{av}}( \mathbf{r},\Delta x) \rangle_{\Delta x>w(\mathbf{r})} = I_M(\mathbf{r}) + I_N ,
\end{equation}
where $\langle \cdots \rangle_{\Delta x> w(\mathbf{r})}$ indicates an average over off-diagonal elements which obey $\Delta x > w(\mathbf{r})$. This average constitutes an average over several realizations of disorder, which is necessary to suppress the fluctuations from constructive and destructive interference between the various possible multiple scattering paths through the sample. 	
Figure \ref{MSfig}(a) shows three examples of normalized intensity profiles $I_{\text{av}}(\mathbf{r},\Delta x)/I_{\text{av}}(\mathbf{r},\Delta x=0)$. Each profile has been averaged over a different zone of the ultrasound image [Fig.~\ref{MSfig}(c)]: green and blue curves (solid and dotted rectangles) correspond to zones situated respectively above and below the bright speckle disk. It is clear that the incoherent background $I_\text{off}$ is higher in the deeper (blue) zone, suggesting that multiple scattering is greatly enhanced behind the reflective object. Surprisingly, the incoherent background $I_\text{off}$ is far from being negligible in the red zone at shallower depths (dashed line rectangle).

To investigate these phenomena further, we define two new observables: (1) the multiple-to-single scattering ratio
 \begin{equation}
 \label{gamma_defeq}
 \rho(\mathbf{r}) \equiv \frac{I_M}{I_S},
 \end{equation}
 and (2) the {multiple scattering-to-noise} ratio,
 \begin{equation}
 \label{epsilon_defeq}
 \epsilon(\mathbf{r}) \equiv \frac{I_M}{I_N}.
 \end{equation}	
To calculate these quantities, it is necessary to be able to distinguish between $I_M$, $I_N$, and $I_S$. 

\subsection{Coherent backscattering as a direct probe of spatial reciprocity}

Discrimination between $I_M$ and $I_N$ can be achieved by exploiting the spatial reciprocity of propagating waves in a linear medium. While the multiple scattering contribution gives rise to a random but symmetric reflection matrix $\Rxx$~\cite{Badon2016,Blondel2018}, additive noise is fully random. Thus, the symmetry of $\Rxx$ gives us a tool to determine the relative weight between noise and multiple scattering in the incoherent background of $\Rxx$. 

An elegant approach to probe spatial reciprocity is the measurement of the CBS effect in the plane-wave basis (the far-field). The CBS effect can be observed by measuring the average backscattered intensity as a function of the angle $\Delta\theta\equiv\left|\theta_\text{in}-\theta_\text{out}\right|$ between the incident and reflected waves. In the presence of multiple scattering, this profile displays a flat plateau (the incoherent background), on top of which sits a CBS cone centered around the exact backscattering angle $\Delta\theta=0$. The cone is solely due to constructive interference from waves following reciprocal paths inside the sample [Fig.\ \ref{MSfig}(b), inset]. Thus, CBS in the far-field is a direct probe of spatial reciprocity in the focused basis~\cite{Aubry2007b,Aubry2008}. 

Appendix \ref{appD} describes how to eliminate the single scattering contribution and extract a far-field intensity profile $ I(\mathbf{r},\Delta\theta) $ for the area $\mathcal{A}$ surrounding each focusing point $\mathbf{r}$. In Fig.\ \ref{MSfig}(b), normalized intensity profiles $I_{\text{av}}( \mathbf{r},\Delta \theta)/I_{\text{av}}(\mathbf{r},\Delta \theta=0)$ are shown for the three areas $\mathcal{A}$ highlighted in Fig.\ \ref{MSfig}(c). For each area, a CBS cone is clearly visible, showing that the experimental data contain contributions from multiple scattering. Just as with the CBS peak in the focused basis [Fig.\ \ref{MSfig}(a)], the highest amount of multiple scattering is observed for the red zone at shallow depths.

	To estimate the relative weight of the noise and multiple scattering contributions, we examine the mean intensity for two cases:  (1) at exact backscattering
	\begin{equation}
		\label{conI}
	I_{\text{av}}(\mathbf{r},\Delta \theta =0)=2I_M(\mathbf{r})+I_N(\mathbf{r}) ,
	\end{equation}
	and (2) at angles away from the CBS peak [Fig.\ \ref{MSfig}(d)]
\begin{equation}
\label{moyI}
\langle I_{\text{av}}(\mathbf{r},\Delta \theta) \rangle_{\Delta \theta > \theta_c} =I_M(\mathbf{r})+I_N(\mathbf{r}) ,
\end{equation}
where $\theta_c$ is the width of the CBS peak and $\langle \cdots \rangle_{\Delta \theta > \theta_c}$ indicates an average over all angles $\Delta\theta$ which obey $\Delta \theta >\theta_c$. 
The enhancement factor of the CBS peak is given by
\begin{equation}
\label{chi}
\chi(\mathbf{r}) = \frac{I_{\text{av}}(\mathbf{r},\Delta \theta=0 )}{\langle  I_{\text{av}}(\mathbf{r},\Delta \theta) \rangle_{\Delta \theta > \theta_c}} .
\end{equation}
$\chi(\mathbf{r'})$ can have values ranging from $1$ to $2$; it is at a minimum when $I_M=0$ and a maximum when all backscattered signal originates from multiple scattering.

\subsection{Maps of multiple scattering rates}
\label{sec_MSmaps}

The multiple scattering-to-noise ratio $\epsilon(\mathbf{r})$ [Eq.~\ref{epsilon_defeq}] can be expressed as a function of the enhancement factor $\chi(\mathbf{r})$ by injecting Eqs.~\ref{conI} and \ref{moyI} into Eq.~\ref{chi} :
\begin{equation}
\epsilon(\mathbf{r}) =\frac{\chi(\mathbf{r})-1 }{2-\chi(\mathbf{r}) } .
\end{equation}
The multiple-to-single scattering ratio $\rho(\mathbf{r})$ [Eq.~\ref{gamma_defeq}] can be derived by injecting the last equation into Eqs.~\ref{Ion} and \ref{Ioff}
\begin{equation}
\rho(\mathbf{r}) = \frac{\left[\chi(\mathbf{r})-1\right] \cdot 	I_{\text{off}}(\mathbf{r}) }{I_{\text{on}}(\mathbf{r})-\chi(\mathbf{r})\cdot 	I_{\text{off}}(\mathbf{r}) } .
\end{equation}

Figures\ \ref{MSfig} (d) and (e) show experimental results for $\rho(\mathbf{r})$ and $\epsilon(\mathbf{r})$, respectively, superimposed onto the original ultrasound image.  
Both of these maps constitute new contrasts which are complementary to the reflectivity maps produced by conventional ultrasound imaging. To begin with, $\rho(\mathbf{r})$ can be used as an indicator of the reliability of a reflectivity image [such as that in Fig.\ \ref{MSfig}(c)]. Because the single-to-multiple scattering ratio is a direct indicator of the validity of the single-scattering (Born) approximation, the reliability of the ultrasound image should scale as the inverse of $\rho(\mathbf{r'})$. An interesting example is displayed in Fig.\ \ref{MSfig}(d) at a depth of $37$~mm (white solid rounded rectangle), where a high multiple-to-single scattering rate is observed. This abrupt increase of\ $\rho(\mathbf{r})$ can be accounted for by double reflection events between the probe and the tissue-phantom interface. We can thus conclude that the structures that seem to emerge in Fig.\ \ref{MSfig}(c) at the same depth are in fact artifacts due to multiple reflections. 

With respect to the quantification of multiple scattering, the parameter $\epsilon(\mathbf{r})$ seems to be particularly relevant. The areas in Fig.\ \ref{MSfig}(e) highlighted by dashed lines exhibit a strong and extended multiple scattering background. \laura{While deeper speckle regions with low scatterer density exhibit a low scattering rate, the bright speckle area in the phantom ($z \sim 60$ mm, white dashed-dotted circle) contains a sufficient concentration of scatterers to generate multiple scattering events. At shallower depths, high amounts of multiple scattering} can be attributed to several small structures indicated by white arrows in Fig.\ \ref{MSfig}(c): (i) two regions in the bovine tissue layer contain air bubbles that generate resonant scattering, thereby inducing a strong multiple scattering `tail' behind them; (ii) a set of bright targets close to each other give rise to strong multiply-scattered echoes at depth $z=$50 mm.  
Figure \ref{MSfig}(e) thus demonstrates how the parameter $\epsilon (\mathbf{r})$ can provide a highly contrasted map of the multiple scattering rate \laura{-- a quantity which is directly related to the density of scatterers~\cite{Sheng2006}}.

\laura{Finally, use of both maps for the interpretation of specific regions can be instructive. While it could be suggested that the enhancement of $\rho(\mathbf{r})$ \alex{[Fig.~\ref{MSfig}(d)]} at the upper dotted lines is due to acoustic shadowing (a decrease in $I_S$\alex{)}, $\epsilon(\mathbf{r})$ indicates that the level of multiple scattering dominates above noise \alex{[Fig.~\ref{MSfig}(e)]}. Thus, one can conclude that multiple scattering is truly increased in this area. \alex{On the contrary, the area on the top left of the phantom shows a large increase of $\rho(\mathbf{r})$ \alex{[Fig.~\ref{MSfig}(d)]} but a weak multiple scattering-to-noise ratio $\epsilon(\mathbf{r})$ \alex{[Fig.~\ref{MSfig}(e)]}. Hence, the high value of $\rho(\mathbf{r})$ is here induced by the acoustic shadow of the bovine tissue layer upstream.} } 

An important technical note is that the study of spatial reciprocity in the reflection matrix requires in principle that the bases of reception and emission be identical. Because this is not the case for our experimental measurements, this tends to slightly underestimate $I_M$.
Relatedly, here we have employed the CBS effect to probe spatial reciprocity, but an equivalent measurement can be performed directly in the focused basis by computing correlations between symmetric elements of $\Rxx$. \\ 

\subsection{Discussion} 

The maps of\ $\rho(\mathbf{r})$ and\ $\epsilon(\mathbf{r})$ help to provide an overall assessment of the factors impacting image quality. For instance, they can be used to explain the apparent poor focus quality in some areas of Fig.\ \ref{ResMap_fig}(b), which appear even after correction for wavefront distortion. The compensation for an incorrect hypothesis for $c$ does not compensate for the effect of multiple scattering or reflections [e.g. in the highlighted areas of Fig.\ \ref{MSfig}(e)]. Thus, the multiple scattering rates combined with the measurement of\ $F(\mathbf{r})$ provide a sensitive local mapping of the heterogeneities in the medium which includes both small- and large-scale variations of the refractive index. In this context, we would like to mention the recent work of Velichko~\cite{Velichko2019}, which measures a quantity similar to $\rho(\mathbf{r})$ as a function of depth and frequency. While noise is not treated separately from multiple scattering, their results emphasize the clear relation between local measurements of multiple scattering and the reliability of ultrasound images. Integration of our focused reflection matrix approach [Eq.\ (\ref{RxxMatrix})] could improve their axial resolution, and help extend their analysis to 2D spatial mapping.

Beyond image reliability, the maps displayed in Fig.\ \ref{MSfig}(d,e) provide a great deal of quantitative information about the system under investigation. Because\ $\epsilon(\mathbf{r})$ is calculated from the off-diagonal elements of\ $\Rxx$, it contains only negligible contributions from single scattering, and thus constitutes an interesting new contrast for imaging which is much more sensitive to the microstructure of the medium than it is to its reflectivity. Conversely, because\ $\rho(\mathbf{r})$ is independent of noise, it can constitute a useful biomarker for medical imaging deep inside tissue, and a potentially valuable tool for future research seeking to characterize multiple scattering media, even at large depths where\ $I_N$ becomes important. 

The perspective of this work will be to extract from\ $\rho(\mathbf{r})$ quantitative maps of scattering parameters such as the elastic mean free path or the absorption length~\cite{Shajahan2014b,Aubry2011}, and transport parameters such as the transport mean free path~\cite{Bayer1993a,Jonckheere2000,Wolf1988c} or the diffusion coefficient~\cite{Bayer1993a,Tourin1997,Cobus2017}. While diffuse tomography in transmission only provides a macroscopic measurement of such parameters, preliminary studies have demonstrated how a reflection matrix recorded at the surface can provide transverse measurements of transport parameters~\cite{Aubry2007b,Aubry2008,Badon2016b,Mohanty2017}. The focused reflection matrix we have introduced here connects each point inside the medium to all other points. Hence, a 2D or 3D map of transport parameters can now be built by solving the radiative transfer inverse problem. \\

\begin{figure*}[t]
\centering
\includegraphics[width=\textwidth]{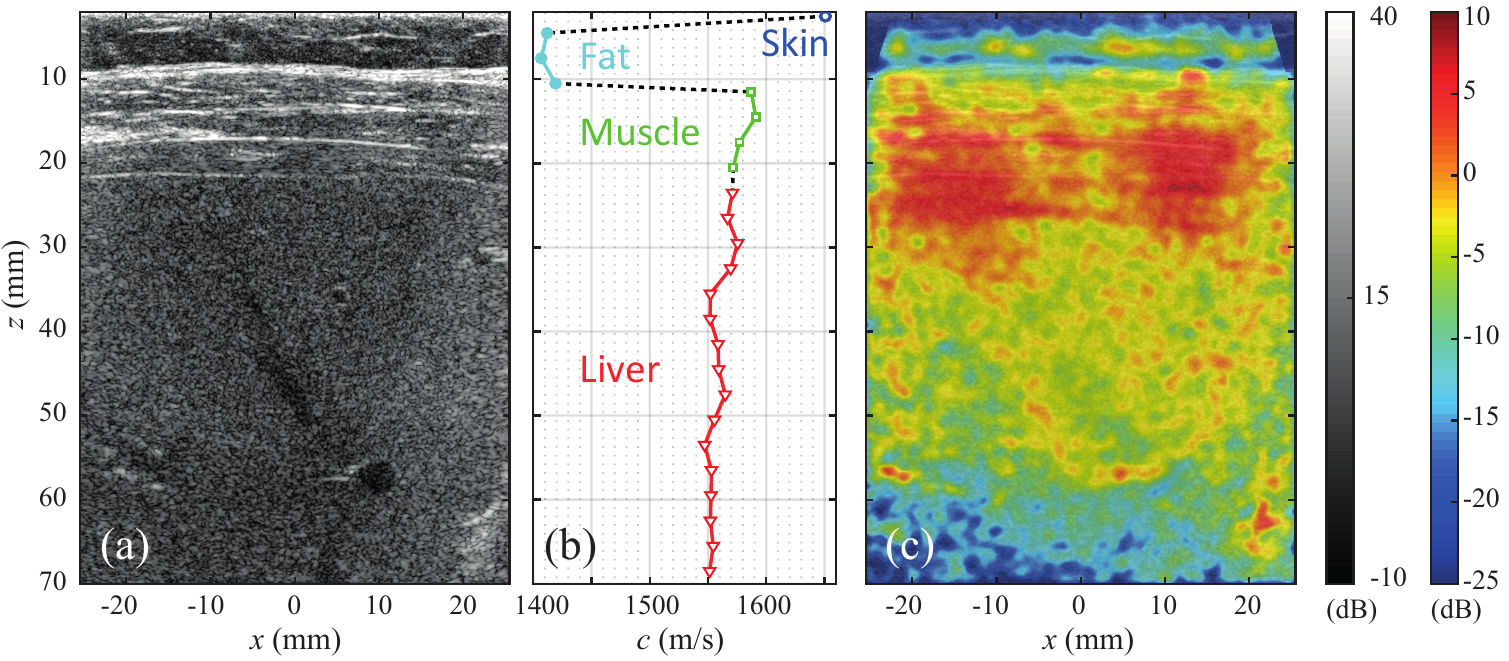}
\caption{In vivo imaging of the human liver using the focused reflection matrix approach. (a) Ultrasound image calculated using a four-layer model [Eq.\ (\ref{imcalc})]. The four layers can be identified as skin, fat, muscle and liver tissue. (b) The speed of sound is calculated for each depth via optimization of $F$ (Section\ \ref{sec_WaveVelocityMapping}). (c) The ratio quantity $\epsilon(\mathbf{r})$ is shown in the same region of interest as (a).}
\label{LiverFig}
\end{figure*}

\section{In vivo quantitative imaging of human tissue}
\label{sec_InvivoLiverImaging}

In this section, we use the focused reflection matrix approach for in vivo quantitative imaging of human tissue. Whereas conventional ultrasound is mostly qualitative, producing images to be analysed by eye, qualitative ultrasound imaging aims to provide numbers which are directly related to the properties of tissue and structures in the body, with the goal of providing information complementary to that of the ultrasound image. As previously discussed, both the speed of sound $c$ and the characteristics of acoustic multiple scattering can be directly related to tissue properties: indeed, there currently exist techniques which use these measurements for qualitative imaging. However, current measurements are greatly limited in terms of spatial resolution, while our approach enables well-resolved maps of $c$ and multiple scattering. Here, we present such maps of the human abdomen and discuss the perspectives for quantitative ultrasound imaging.

The reflection matrix $\Rut$ was acquired with the probe in contact with the abdomen of a healthy volunteer. The ultrasound sequence is the same as the one use for the in vitro phantom study of Sections\ \ref{sec_LocalFocusingCriterion} and\ \ref{sec_MultipleScattering}. The study was performed in conformation with the declaration of Helsinki. The resulting ultrasound image (computed using a four-layer model) is shown in Fig.\ \ref{LiverFig}(a). Quantitative imaging maps were calculated in post-processing -- computational details are discussed in Appendix\ \ref{appC}.

Figure\ \ref{LiverFig}(b) shows the speed of sound plotted as a function of depth. From this plot, four distinct tissue layers can be identified: skin, fat, muscle, and liver tissue. We are thus able to estimate $c$ for each tissue. In the skin, previous authors have reported speed of sound values in the range of $c_\mathrm{skin}\approx 1500 - 1750$~m/s, with an average value of $c_\mathrm{skin}\sim 1625$~\cite{Moran1995}. The wide range of values for $c_\mathrm{skin}$ is most likely due to \laura{the significant sensitivity of this parameter on skin hydration, as well as variations in temperature, age of the cadaver skin examined, and region of the body from which the skin was extracted.} Thus, more accurate approaches for this measurement would be valuable. Our method gives an estimate of $c_\mathrm{skin}\approx 1651$~m/s, which to our knowledge constitutes the first in vivo measurement of $c_\mathrm{skin}$ in this frequency range. In the fat layer, we find an average value of $c_\mathrm{fat}= 1413\pm 6$~m/s. (The standard deviation of the values in this layer is used as an estimate of the experimental uncertainty.) Our result agrees with previously reported results of $c_\mathrm{fat}= 1427\pm 12.7$~m/s~\cite{Errabolu1998}. In the muscle layer, our measured average value of $c_\mathrm{muscle}=1582\pm 9$~m/s agrees with the commonly cited value of $c_\mathrm{muscle}=1576\pm 1.1$~m/s~\cite{Rajagopalan1979}. Finally, we find an average speed of sound in the liver of $c_\mathrm{liver}=1559\pm 8$~m/s, consistent with previous measurements in healthy human liver~\cite{Lin1987,Boozari2010,Imbault2017,Stahli2019}. Overall, this approach enables the simultaneous measurement of $c$ in four human tissue layers using one experimental data set, with no dependence on the initial guess for $c$. It thus constitutes a significant advance over state of the art methods for $c$ measurement in human tissue (c.f. Refs~\cite{Imbault2017,Jakovljevic2018,Stahli2019}). \\

Figure\ \ref{LiverFig}(c) shows a spatial map of $\epsilon(\mathbf{r})$, the ratio of multiple scattering to noise. Discrete areas in which $\epsilon(\mathbf{r})$ is very high ($>0$~dB) are indicative of artifacts caused by multiple reverberations from the tissue layers ($z=10-30$~mm) or in structures such as veins, for instance at $(z,x)\approx(56,5)$~mm. 
Strikingly, we also find a significant amount of multiple scattering relatively evenly distributed across areas of speckle (compared to the phantom in Fig.\ \ref{MSfig}).

Acoustic multiple scattering is extremely sensitive to tissue microstructure, and thus can be a useful indicator of tissue health. Based on this concept, there are current methods which attempt to quantify observables of acoustic scattering.
Statistical parameters measured from the backscatter coefficient (BSC) of ultrasonic speckle [a quantity related to our $\rho(\mathbf{r})$] can give estimates of scatterer size and density~\cite{Francechini2019,Oelze2016}. However, BSC measures the entire backscattered energy, and thus does not distinguish between multiple and single scattering. Our measurement of $\epsilon(\mathbf{r})$, on the other hand,  provides two dimensional, well-resolved, spatial maps of the rate of multiple scattering inside human tissue with a negligible dependence on tissue echogeneity. It is thus truly a new quantitative imaging contrast. Previous works have shown that the rate of acoustic multiple scattering can be used to distinguish between healthy and unhealthy tissue~\cite{Mohanty2018,Aubry2011}; thus, $\epsilon(\mathbf{r})$ and $\rho(\mathbf{r})$ present an important advance for qualitative ultrasound imaging. Our mapping approach could also be used to help increase the spatial resolution of statistical analyses such as those based on the BSC. Our approach also offers a significant advantage for assessing image reliability. In recent work, a coherence-based approach was used to calculate acoustic multiple scattering and thermal noise for an image quality metric~\cite{Long2018}; however, the two contributions are not separated, and an average must be taken over relatively large areas of the region of interest. By separating and contrasting these contributions, our method is sensitive to the origin of a decrease in reliability (thermal or acoustic noise), thus giving a more informed picture of the system under study. Moreover, our double focusing approach and the subsequent common mid-point analysis should provide, in principle, a better spatial resolution.  \\

Finally, we discuss the experimental limitations of the methods presented in this section. The measurement of $c(z)$ is limited primarily by the depth from which singly scattered echoes can be detected. The speed of sound estimation is based on the minimization of aberrations undergone by singly-scattered signals. The major physical limitation is thus linked to the amount of singly-scattered signals detected. For imaging through bones or through air, this ratio is most highly impacted by attenuation; thus, the depth limitations here are similar to those for other conventional imaging techniques (the ratio of singly-scattered signals to noise). For imaging in lungs, bone, and other highly scattering media, depth is most strongly limited by the ratio of singly- to multiply-scattered signals. Deeper than one transport mean free path $\ell^*$, signals become completely randomized by multiple scattering, and no singly-scattered signals will be measurable. Weakly scattering tissue can be characterized as that in which singly-scattered signals exist, but in which multiple scattering significantly degrades the quality of conventional ultrasound images. In these tissues, such as breast or muscle, our measurements of $c$ should still be effective~\cite{Aubry2011}. In more highly scattering tissue, however, our current method may not be of much use: for example, values of $\ell^*\approx 0.3-1.1$~mm have been measured at 8~MHz in lung tissue~\cite{Mohanty2017}.

On the other hand, it will be interesting (and is immediately possible) to create maps of $\rho(\mathbf{r})$ and $\epsilon(\mathbf{r})$ in scattering media such as breast, lung and bone. Recent work such as that by Mohanty et al.~\cite{Mohanty2018} suggests that such maps may be better at imaging heterogeneous scattering media than conventional echographic ultrasound. However, for bone and flat layers of tissue such as muscle, a current limitation is the coexistence of multiple scattering and artifacts \laura{from reverberant echos or reflections caused by interfaces \alex{between} tissues with different \alex{acoustic} impedances. The separation of these effects will be the subject of future work.} 

\section{Conclusion and perspectives}
\label{conclusion}

In summary, a powerful and elegant matrix approach for quantitative wave imaging has been presented. By focusing at distinct points in emission and reception, one can build a focused reflection matrix that contains the impulse responses between a set of virtual transducers mapping the entire medium. From this focused reflection matrix, a local focusing parameter can be estimated at any point of the inspected medium. 
Because it can be applied to any type of media, including in the speckle regime or in the presence of specular reflectors, this focusing criterion is suitable to any situation encountered in medical ultrasound, and enables wave velocity mapping of the medium.  In this paper, we have demonstrated proofs of concept for this approach using a two-layer phantom system, as well as in vivo measurements on the human abdomen in which $c$ was simultaneously measured for four separate tissue layers. This physically intuitive approach does not depend on arbitrary parameters such as image quality or the initial guess for $c$, and does not require guide stars or complex iterative adaptive focusing schemes.
Knowledge of the spatial variation of velocity can in turn be used with the focused reflection matrix to overcome wavefront distortions. The contrast and resolution of the image could then be restored almost as if the inhomogeneities had disappeared. We have shown not only how to measure $c(z)$, but how it can then be applied to overcome phase aberration using the same experimental data set. Importantly, this method has the potential to treat spatially-varying aberrations: this perspective will be the subject of future works.

We have also shown that the focused reflection matrix enables a local examination of multiple scattering processes deep inside the medium. We have demonstrated the effectiveness of using fundamental interference phenomena such as coherent backscattering -- a hallmark of multiple scattering processes -- to discriminate between multiple scattering and measurement noise. A novel imaging method is proposed based on the multiple scattering contrast. To our knowledge, such 2D maps have never before been demonstrated, and current state-of-the-art methods can not produce such well-resolved local information about acoustic multiple scattering. Unexplored but promising perspectives for this work include the quantitative imaging of parameters such as the scattering, absorption or transport mean free paths. 

\alex{One limit of the reflection matrix approach lies on the linear theory on which it relies\lauraNEW{, meaning that it is inherently incapable of accounting} for nonlinear effects. Nevertheless, \lauraNEW{in} our opinion, the focused reflection matrix can still be of interest for optimizing tissue harmonic imaging~\cite{Whittingham}. First, aberration correction or a better wave velocity model can lead to an optimized input focusing process and a more efficient nonlinear conversion at the focus. More generally, a nonlinear reflection matrix linking, for instance, input focusing points at the fundamental frequency and output focusing points at the harmonic frequency may be a useful tool to optimize the non-linear conversion process. A matrix approach of harmonic imaging \lauraNEW{will} be the subject of future works.}

Finally, we emphasize that we have only concentrated here on the relationship between virtual transducers located in the same focal planes at the ballistic time. This has been applied to a medium which can be modeled by a stack of various horizontal layers. It is equally possible to consider responses between, for example, angled or curved focal planes, which could simplify similar quantitative imaging in organs such as the brain. More generally, there is enormous further potential for the analysis of the entire focused matrix\ $\Rrr$ across the whole medium and beyond the ballistic time, which will be explored in future works. Last but not least, our matrix approach of wave imaging is very general and could be applied to any kind of system for which emission and detection of waves can be varied in a controllable way. 
Thus, the potential of this work goes far beyond ultrasound imaging, with immediate foreseeable impacts in a range of wave physics including optical microscopy, radar and seismology. 

\begin{acknowledgments}
	The authors are grateful for funding provided by LABEX WIFI (Laboratory of Excellence ANR-10-LABX-24) within the French Program “Investments for the Future” under Reference No. ANR-10-IDEX-0001-02 PSL*. This project has received funding from the European Research Council (ERC) under the European Union’s Horizon 2020 research and innovation programme (grant agreement No. 819261). W.L. acknowledges financial support from the SuperSonic Imagine company. L.C. acknowledges financial support from the European Union’s Horizon 2020 research and innovation programme under the Marie Sk\l{}odowska-Curie grant agreement No. 744840. 
\end{acknowledgments}

\appendix

\section{\label{appA}Experimental procedure}

The experimental set up consisted in an ultrasound phased-array probe (SuperLinear$^{\text{TM}}$ SL15-4, Supersonic Imagine) connected to an ultrafast scanner (Aixplorer\textregistered, SuperSonic Imagine, Aix-en-provence, France). This 1D array of 256 transducers with a pitch $p=0.2$ mm was used to emit $41$ plane waves with an angle of incidence $\tin$ spanning from\ $-20^o$ to $20^o$ [Fig.\ \ref{acq}(a)]. The emitted signal was a sinusoidal burst of central frequency\ $f_c=7.5$~MHz, with a frequency bandwidth spanning from\ $2.5$ to $10$~MHz. In reception, all elements were used to record the reflected wavefield over a time length $t=124$~$\mu$s at a sampling frequency of 30 MHz. The ultrasound sequence is driven by using the research pack of the Aixplorer device (SonicLab, Supersonic Imagine, France). The matrix acquired in this way is denoted\ $\Rut(t)\equiv R(\uout,\tin,t)$. 

\section{\label{appB}Derivation of the common-midpoint intensity profile}

The theoretical expression of the common-midpoint intensity profile $I(\mathbf{r},\Delta x )$ is derived in the specular and speckle scattering regimes.

In the specular scattering regime, the characteristic size $l_{\gamma}$ of reflectors is much larger than the width of the focal spot $\delta x$. $\gamma(\mathbf{r})$ can thus be assumed as invariant over the input and output focal spots. Equation \ref{Rrr_intermsofh_eq} then becomes
\begin{equation}
\label{RrDr_spec}
R(\mathbf{r},\Delta x )=    \gamma(\mathbf{r}) \times  \left (  H_\text{in}  \ast H_\text{out} \right) (\Delta \mathbf{r}).
\end{equation}
The injection of Eq.\ \ref{RrDr_spec} into Eq.\ \ref{IrDr_fromRrDr} yields the expression of $I(\mathbf{r},\Delta x )$ given in Eq.~\ref{IrDr_spec}.

In the speckle scattering regime, $l_{\gamma}<<\delta x$. This is the most common regime in ultrasound imaging, as scattering is more often due to a random distribution of unresolved scatterers. To a first approximation, such a random medium has the property that
\begin{equation}
\label{random}
\langle \gamma (\mathbf{r_1}) \gamma^*(\mathbf{r_2})\rangle = \langle \left | \gamma \right|^2 \rangle \delta(\mathbf{r_2}-\mathbf{r_1}),
\end{equation}
where $\delta$ is the Dirac distribution. The combination of Eqs.~\ref{Rrr_intermsofh_eq}, \ref{IrDr_fromRrDr} and \ref{random} gives directly $I(\mathbf{r},\Delta x )$ as expressed in Eq.~\ref{IrDr_fromRrDr2}.

\section{\label{appC} Computational details}

All calculations shown in this paper were performed in Matlab. To quantify the resources required for these computations, we can compare the time required for our quantitative matrix approach with that required to create a plane-wave compounded ultrasound image~\cite{Montaldo2009}. For a single plane-wave transmission, the imaging time $t_1$ combines the time required to record the reflected wave-field and that to focus at reception (with software) on the $N_x\times N_z$ points of the image. The acquisition of 70 mm depth images can be typically produced for a time $t_1\sim100$ $\mu$s ($\sim$10 kfps~\cite{Tiran_2015}). To estimate the common mid-point intensity profile over a distance $\Delta x_\textrm{max}$, we need to record $N_{\Delta x}=\Delta x_\textrm{max}/\delta x_0$ sub-diagonals of the focused reflection matrix $\mathbf{R}_{xx}(z)$ [see Eq.\ (\ref{IrDr_fromRrDr}) and the accompanying text]. To retrieve these sub-diagonals, the reflection matrix $\mathbf{R}_{\mathbf{r}\theta}$ should be initially recorded with a set of $N_{\Delta x}$ plane-waves. In our case, $N_{\Delta x}=41$ (see Appendix A). Those recorded wave-fields shall then be focused at reception and recombined at emission to form the focused reflection matrices. The time $t_M$ for getting the set of matrices $\mathbf{R}_{xx}(z)$ is thus $t_M=N_{\Delta x}\times t_1 \sim 4$ ms ($\sim$ 250 fps). Once this set of matrices has been synthesized, the multiple scattering analysis is straight-forward. Maps of the multiple scattering rates, $\rho(\mathbf{r})$ and $\rho(\mathbf{r})$, can thus be easily obtained in real-time ($>$ 25 fps) since it  approximately requires the same number $N_{\Delta x}$ of transmits as that required to build the compounded image. If needed, this time can be greatly decreased by selecting only a limited region of interest in which to map $\epsilon(\mathbf{r})$ or  $\rho(\mathbf{r})$.

To perform the optimization over the  wave velocity $c$, we need to perform the aforementioned operations for a number $N_c$ of test values of $c$. The time required for the speed of sound mapping is then $t_c=N_c \times t_M = N_c\times N_{\Delta x} \times t_1 \sim 80$ ms ($\sim$ 12 Hz). If we take, for example, a range of $c=100$~m/s with a step of $5$~m/s, then $N_c=20$. To reach real-time imaging, one possibility is to reduce $N_{\Delta x} $ since the single-scattering contribution lies along the near-diagonal coefficients of $\mathbf{R}_{xx}(z)$. Only a few of its sub-diagonals ($N_{\Delta x}\sim 10$) are thus needed to assess the focusing criterion $F(\mathbf{r})$. The range of $c$ can also be narrowed and parallel computing employed to cut down on processing time. Due to these considerations, we expect that this computation could in the near future be performed in real time. 

Of the analyses presented in this paper, only the measurement of $c(z)$ is not fully automated, as it requires the user to identify the approximate regions in which different values of $c$ should be anticipated [i.e. to differentiate between the four different layers in Fig.\ \ref{LiverFig}(a)]. To make the process completely user independent would require more sophisticated coding to perform this image segmentation -- something that is, we believe, feasible in an industrial setting but beyond the scope of this paper. 

\hfill
\section{\label{appD} Measurement errors on the focusing criterion and the speed of sound}

In this work, we have defined the focusing parameter $F$ as the ratio between the width $w_0$ of the ideal diffraction-limited PSF and the width $w$ of the experimentally measured PSF. In optics, the Strehl $S$ ratio is generally used to quantify aberration~\cite{mahajan1982strehl}. It is defined as the ratio between the maximum of the PSF intensity, $I$, and that in the ideal diffraction-limited case, $I_0$. Due to energy conservation, we have $ I_0 \times w_0 = I \times w$. The focusing criterion and Strehl ratio, as well as their relative measurement errors, are thus equivalent:
\begin{equation}
F \equiv S 
\end{equation}
and
\begin{equation}
\label{errorF}
\frac{\Delta F}{F}\equiv \frac{\Delta S}{S}.
\end{equation}
$S$ can also be expressed as the square magnitude of the averaged aberration transmittance $e^{i  \phi(\sin \theta)}$~\cite{mahajan1982strehl}:
\begin{equation}
\label{S}
S=\left | \left \langle  e^{i \phi(\sin \theta)}  \right \rangle_{\sin \theta}\right |^2 ,
\end{equation}
where $ \phi(\sin \theta)$ is the far-field phase delay induced by the mismatch between the propagation model and the real medium in the $\theta$-direction. 

In Fig.\ \ref{qualityoffocus_img}, a two-layer medium is used to model the bovine tissue/phantom system. Assuming that the wave velocity $c_t$ is properly estimated in the first layer (bovine tissue), the phase $ \phi(\sin \theta)$ accumulated in the phantom is given by
\begin{equation}
\phi(\sin \theta)=k_p z_p \cos(\theta_p) ,
\end{equation}
where $k_p=\omega/c_p$ is the wavenumber in the phantom and $\theta_p$ is the refraction angle in the phantom, obeying $\sin \theta_p/c_p=\sin \theta /c_t $. If a wrong value of $c_p$ is used to model sound propagation in the phantom, the resulting phase distortion is given by
\begin{equation}
\Delta \phi(\sin \theta)= - \frac{k_p z_p}{\cos \theta_p }  \eta ,
\end{equation}
where $\eta=\Delta c_p /c_p$ is the relative error
of the speed of sound hypothesis in the phantom. For the sake of simplicity, we will assume in the following that $\cos \theta \sim \cos \theta_p $. This approximation is justified by the small relative difference between $c_p$ and $c_t$. 
Assuming relatively weak aberrations ($ \Delta \phi(\sin \theta) << \pi $), the transmittance aberration function $e^{i \Delta \phi(\sin \theta)}$ can be expanded as 
\begin{eqnarray}
e^{i \Delta \phi(\sin \theta)}  \sim  1- i \frac{{k_p }  z_p}{ \cos \theta}  \eta &-& \frac{1}{2}\left ( \frac{k_p z_p }{\cos \theta} \right )^2  \eta^2 \nonumber \\
 &  &+ \mathcal{O}\left ( \eta^3 \right ) .
\end{eqnarray}
The angular average of $e^{i  \phi(\sin \theta)}$ is then deduced
\begin{eqnarray}
\left \langle e^{i \Delta \phi(\sin \theta)} \right \rangle_{\sin \theta} & = & \frac{1}{\sin \beta} \int_0^{\sin \beta}    e^{i \Delta \phi(\sin\theta)} d(\sin \theta)
\nonumber \\
 & \sim &  1 - i   k_p z_p \eta \frac{ \beta}{\sin \beta} \nonumber \\
 & & - \frac{1}{2} \left ( k_p z_p \eta \right )^2 \frac{\arctanh (\sin \beta)}{\sin \beta} \nonumber  \\
&  &+ \mathcal{O}\left ( \eta^3 \right ) .
\end{eqnarray}
Injecting the last expression into Eq.\ \ref{S} leads to the following expression of the Strehl ratio  :
\begin{eqnarray}
S  \sim  1  &- &  \frac{(k_p z_p \eta)^2   }{\sin \beta}\left [ \arctanh (\sin \beta) - \frac{ \beta^2}{\sin \beta}\right ] \nonumber \\
& + &\mathcal{O}\left ( \eta^3 \right ) .
\end{eqnarray}
For weak aberrations ($F,S\sim 1$), the relative error $\Delta F/F$ [Eq.~\ref{errorF}] of the focusing criterion can then be directly deduced from the previous expansion of the Strehl ratio:
\begin{equation}
\label{errorF_appendix}
\frac{\Delta F}{F} = \frac{(k_p z_p \eta)^2}{\sin \beta }   \left [ \arctanh (\sin \beta) - \frac{ \beta^2}{\sin \beta}\right ] .
\end{equation}

\section{\label{appE}Computation of local intensity profiles in the plane wave basis}

To quantify the CBS effect we first need to eliminate contributions from single scattering. To this end, the reflection matrices\ $\Rxx(z)$ are first normalized such that their diagonal at each depth exhibits a constant mean intensity:
\begin{equation}
\bar{R}(x-\Delta x,x+\Delta x,z)=\frac{R(x-\Delta x,x+\Delta x,z)}{\sqrt{  I_{\text{av}}(\mathbf{r},\Delta x) }} .
\end{equation}
This operation eliminates the dominant contribution to intensity from diagonal elements in $\Rxx$, which is equivalent to removal of the single scattering component. The matrix formalism makes it easy to then project $\mathbf{ \bar{R}_{xx}}(z)$ into the plane-wave basis. The matrix approach also means that it is simple to project only a subspace of $\mathbf{ \bar{R}_{xx}}(z)$ into the plane-wave basis. For each point $\mathbf{r}$ of the image, we define a sub-space matrix $\mathbf{M}_{xx}(z,\mathbf{r})$ whose non-zero coefficients ${M}(\xout,\xin,z,\mathbf{r}) $ are associated with common midpoints $\mathbf{r'}=(\rin+\rout)/2 $ belonging to the area $\mathcal{A}$ surrounding $\mathbf{r}$:
\begin{equation*}
  M(\xout,\xin,z,\mathbf{r})  = \left \{
\begin{array}{ll}
\bar{R}(\xout,\xin,z) & \, \mbox{for }(\mathbf{r'}-\mathbf{r}) \in \mathcal{A} \\
0 & \,  \mbox{elsewhere}  .
\end{array} 
\right. 
\end{equation*}
With this set of sub-matrices $\mathbf{M}_{xx}(z,\mathbf{r})$, one can locally probe the far-field CBS. Projection of $\mathbf{M}_{xx}(z,\mathbf{r})$ into the plane-wave basis is performed at each depth using the transmission matrices $\mathbf{T_0}(z,\omega_c)$  [Eq.~\ref{Gzp_eq}]:
\begin{equation*}
\label{projRtt}
\mathbf{ M_{\theta \theta}}(z,\mathbf{r})=\mathbf{T_0^{\top}}\left(z,\omega_c\right)\times \mathbf{M}_{xx}(z,\mathbf{r}) \times\mathbf{T_0}\left(z,\omega_c\right)  .
\end{equation*}
$\mathbf{M}_{\theta \theta}(z,\mathbf{r})$ contains the normalized reflection coefficients in the $\theta_\text{out}$ direction for an angle of incidence $\theta_\text{in}$ induced by scatterers contained in the area $\mathcal{A}$ centered around $\mathbf{r}$. An averaged far-field mean intensity can now be calculated as a function of the reflection angle $\Delta \theta$:
\begin{equation*}
\label{eqmsI}
I_{\text{av}}(\mathbf{r},\Delta \theta)=\left \langle \left | M(\theta + \Delta \theta /2,\mathbf{\theta}-\Delta \theta /2,z,\mathbf{r}) \right |^2 \right \rangle_{\theta,z} ,
\end{equation*}
where the symbol $\langle \cdots \rangle$ denotes an average over the variables in the subscript, i.e. all angles which obey $\theta=(\theta_\text{in}+\theta_\text{out})/2$ and the thickness of the area $\mathcal{A}$.
%


\begin{thebibliography}{84}%
\makeatletter
\providecommand \@ifxundefined [1]{%
 \@ifx{#1\undefined}
}%
\providecommand \@ifnum [1]{%
 \ifnum #1\expandafter \@firstoftwo
 \else \expandafter \@secondoftwo
 \fi
}%
\providecommand \@ifx [1]{%
 \ifx #1\expandafter \@firstoftwo
 \else \expandafter \@secondoftwo
 \fi
}%
\providecommand \natexlab [1]{#1}%
\providecommand \enquote  [1]{``#1''}%
\providecommand \bibnamefont  [1]{#1}%
\providecommand \bibfnamefont [1]{#1}%
\providecommand \citenamefont [1]{#1}%
\providecommand \href@noop [0]{\@secondoftwo}%
\providecommand \href [0]{\begingroup \@sanitize@url \@href}%
\providecommand \@href[1]{\@@startlink{#1}\@@href}%
\providecommand \@@href[1]{\endgroup#1\@@endlink}%
\providecommand \@sanitize@url [0]{\catcode `\\12\catcode `\$12\catcode
  `\&12\catcode `\#12\catcode `\^12\catcode `\_12\catcode `\%12\relax}%
\providecommand \@@startlink[1]{}%
\providecommand \@@endlink[0]{}%
\providecommand \url  [0]{\begingroup\@sanitize@url \@url }%
\providecommand \@url [1]{\endgroup\@href {#1}{\urlprefix }}%
\providecommand \urlprefix  [0]{URL }%
\providecommand \Eprint [0]{\href }%
\providecommand \doibase [0]{https://doi.org/}%
\providecommand \selectlanguage [0]{\@gobble}%
\providecommand \bibinfo  [0]{\@secondoftwo}%
\providecommand \bibfield  [0]{\@secondoftwo}%
\providecommand \translation [1]{[#1]}%
\providecommand \BibitemOpen [0]{}%
\providecommand \bibitemStop [0]{}%
\providecommand \bibitemNoStop [0]{.\EOS\space}%
\providecommand \EOS [0]{\spacefactor3000\relax}%
\providecommand \BibitemShut  [1]{\csname bibitem#1\endcsname}%
\let\auto@bib@innerbib\@empty
\bibitem [{\citenamefont {Szabo}(2004)}]{Szabo2004}%
  \BibitemOpen
  \bibfield  {author} {\bibinfo {author} {\bibfnamefont {T.~L.}\ \bibnamefont
  {Szabo}},\ }\href@noop {} {\emph {\bibinfo {title} {Diagnostic ultrasound
  imaging: inside out}}}\ (\bibinfo  {publisher} {Elsevier Academic Press, San
  Diego, CA},\ \bibinfo {year} {2004})\BibitemShut {NoStop}%
\bibitem [{\citenamefont {Drexler}\ and\ \citenamefont
  {Fujimoto}(2008)}]{Drexler2008}%
  \BibitemOpen
  \bibfield  {author} {\bibinfo {author} {\bibfnamefont {W.}~\bibnamefont
  {Drexler}}\ and\ \bibinfo {author} {\bibfnamefont {J.~G.}\ \bibnamefont
  {Fujimoto}},\ }\href@noop {} {\emph {\bibinfo {title} {Optical coherence
  tomography}}}\ (\bibinfo  {publisher} {Springer-Verlag, Berlin},\ \bibinfo
  {year} {2008})\BibitemShut {NoStop}%
\bibitem [{\citenamefont {Bamler}\ and\ \citenamefont
  {Hartl}(1998)}]{Bamler1998}%
  \BibitemOpen
  \bibfield  {author} {\bibinfo {author} {\bibfnamefont {R.}~\bibnamefont
  {Bamler}}\ and\ \bibinfo {author} {\bibfnamefont {P.}~\bibnamefont {Hartl}},\
  }\bibfield  {title} {\bibinfo {title} {Synthetic aperture radar
  interferometry},\ }\href@noop {} {\bibfield  {journal} {\bibinfo  {journal}
  {Inverse Problems}\ }\textbf {\bibinfo {volume} {14}},\ \bibinfo {pages} {R1}
  (\bibinfo {year} {1998})}\BibitemShut {NoStop}%
\bibitem [{\citenamefont {Yilmaz}(2008)}]{Yilmaz2008}%
  \BibitemOpen
  \bibfield  {author} {\bibinfo {author} {\bibfnamefont {O.}~\bibnamefont
  {Yilmaz}},\ }\href@noop {} {\emph {\bibinfo {title} {Seismic data analysis.
  Processing, Inversion, and Interpretation of seismic data}}}\ (\bibinfo
  {publisher} {Society of Exploration Geophysicists,Tulsa, OK, USA},\ \bibinfo
  {year} {2008})\BibitemShut {NoStop}%
\bibitem [{\citenamefont {Roddier}(1999)}]{Roddier1999}%
  \BibitemOpen
  \bibfield  {author} {\bibinfo {author} {\bibfnamefont {F.}~\bibnamefont
  {Roddier}},\ }\href@noop {} {\emph {\bibinfo {title} {Adaptive optics in
  astronomy}}}\ (\bibinfo  {publisher} {Cambridge University Press,
  Cambridge},\ \bibinfo {year} {1999})\BibitemShut {NoStop}%
\bibitem [{\citenamefont {Booth}(2007)}]{Booth}%
  \BibitemOpen
  \bibfield  {author} {\bibinfo {author} {\bibfnamefont {M.~J.}\ \bibnamefont
  {Booth}},\ }\bibfield  {title} {\bibinfo {title} {Adaptive optics in
  microscopy},\ }\href@noop {} {\bibfield  {journal} {\bibinfo  {journal}
  {Philos. Trans. R. Soc. A}\ }\textbf {\bibinfo {volume} {365}},\ \bibinfo
  {pages} {2829 } (\bibinfo {year} {2007})}\BibitemShut {NoStop}%
\bibitem [{\citenamefont {O'Donnell}\ and\ \citenamefont
  {Flax}(1988)}]{ODonnell1988}%
  \BibitemOpen
  \bibfield  {author} {\bibinfo {author} {\bibfnamefont {M.}~\bibnamefont
  {O'Donnell}}\ and\ \bibinfo {author} {\bibfnamefont {S.}~\bibnamefont
  {Flax}},\ }\bibfield  {title} {\bibinfo {title} {{Phase-aberration correction
  using signals from point reflectors and diffuse scatterers: measurements}},\
  }\href@noop {} {\bibfield  {journal} {\bibinfo  {journal} {IEEE Trans.
  Ultrason., Ferroelectr., Freq. Control}\ }\textbf {\bibinfo {volume} {35}},\
  \bibinfo {pages} {768} (\bibinfo {year} {1988})}\BibitemShut {NoStop}%
\bibitem [{\citenamefont {Mallart}\ and\ \citenamefont
  {Fink}(1994)}]{Mallart1994}%
  \BibitemOpen
  \bibfield  {author} {\bibinfo {author} {\bibfnamefont {R.}~\bibnamefont
  {Mallart}}\ and\ \bibinfo {author} {\bibfnamefont {M.}~\bibnamefont {Fink}},\
  }\bibfield  {title} {\bibinfo {title} {{Adaptive focusing in scattering media
  through sound-speed inhomogeneities: The van Cittert Zernike approach and
  focusing criterion}},\ }\href@noop {} {\bibfield  {journal} {\bibinfo
  {journal} {J. Acoust. Soc. Am.}\ }\textbf {\bibinfo {volume} {96}} (\bibinfo
  {year} {1994})}\BibitemShut {NoStop}%
\bibitem [{\citenamefont {Park}\ \emph {et~al.}(2018)\citenamefont {Park},
  \citenamefont {Depeursinge},\ and\ \citenamefont {Popescu}}]{Park2018}%
  \BibitemOpen
  \bibfield  {author} {\bibinfo {author} {\bibfnamefont {Y.}~\bibnamefont
  {Park}}, \bibinfo {author} {\bibfnamefont {C.}~\bibnamefont {Depeursinge}},\
  and\ \bibinfo {author} {\bibfnamefont {G.}~\bibnamefont {Popescu}},\
  }\bibfield  {title} {\bibinfo {title} {Quantitative phase imaging in
  biomedicine},\ }\href@noop {} {\bibfield  {journal} {\bibinfo  {journal}
  {Nat. Photonics}\ }\textbf {\bibinfo {volume} {12}},\ \bibinfo {pages} {578}
  (\bibinfo {year} {2018})}\BibitemShut {NoStop}%
\bibitem [{\citenamefont {Kak}\ and\ \citenamefont {Slaney}(2001)}]{Kak2001}%
  \BibitemOpen
  \bibfield  {author} {\bibinfo {author} {\bibfnamefont {A.~C.}\ \bibnamefont
  {Kak}}\ and\ \bibinfo {author} {\bibfnamefont {M.}~\bibnamefont {Slaney}},\
  }\href@noop {} {\emph {\bibinfo {title} {Principles of Computerized
  Tomographic Imaging}}}\ (\bibinfo  {publisher} {Society for Industrial and
  AppliedMathematics, Philadelphia, PA},\ \bibinfo {year} {2001})\BibitemShut
  {NoStop}%
\bibitem [{\citenamefont {Ali}\ and\ \citenamefont {Dahl}(2018)}]{Ali2018}%
  \BibitemOpen
  \bibfield  {author} {\bibinfo {author} {\bibfnamefont {R.}~\bibnamefont
  {Ali}}\ and\ \bibinfo {author} {\bibfnamefont {J.}~\bibnamefont {Dahl}},\
  }\bibfield  {title} {\bibinfo {title} {Distributed phase aberration
  correction techniques based on local sound speed estimates},\ }in\ \href@noop
  {} {\emph {\bibinfo {booktitle} {IEEE Int. Ultrason. Symp.}}}\ (\bibinfo
  {year} {2018})\ pp.\ \bibinfo {pages} {1--4}\BibitemShut {NoStop}%
\bibitem [{\citenamefont {Rau}\ \emph {et~al.}(2019)\citenamefont {Rau},
  \citenamefont {Schweizer}, \citenamefont {Vishnevskiy},\ and\ \citenamefont
  {Goksel}}]{Rau2019}%
  \BibitemOpen
  \bibfield  {author} {\bibinfo {author} {\bibfnamefont {R.}~\bibnamefont
  {Rau}}, \bibinfo {author} {\bibfnamefont {D.}~\bibnamefont {Schweizer}},
  \bibinfo {author} {\bibfnamefont {V.}~\bibnamefont {Vishnevskiy}},\ and\
  \bibinfo {author} {\bibfnamefont {O.}~\bibnamefont {Goksel}},\ }\bibfield
  {title} {\bibinfo {title} {Ultrasound aberration correction based on local
  speed-of-sound map estimation},\ }in\ \href@noop {} {\emph {\bibinfo
  {booktitle} {IEEE Int. Ultrason. Symp.}}}\ (\bibinfo  {publisher} {IEEE},\
  \bibinfo {address} {Glasgow},\ \bibinfo {year} {2019})\ pp.\ \bibinfo {pages}
  {2003--2006}\BibitemShut {NoStop}%
\bibitem [{\citenamefont {Chau}\ \emph {et~al.}(2019)\citenamefont {Chau},
  \citenamefont {Jakovljevic}, \citenamefont {Lavarello},\ and\ \citenamefont
  {Dahl}}]{Chau2019}%
  \BibitemOpen
  \bibfield  {author} {\bibinfo {author} {\bibfnamefont {G.}~\bibnamefont
  {Chau}}, \bibinfo {author} {\bibfnamefont {M.}~\bibnamefont {Jakovljevic}},
  \bibinfo {author} {\bibfnamefont {R.}~\bibnamefont {Lavarello}},\ and\
  \bibinfo {author} {\bibfnamefont {J.}~\bibnamefont {Dahl}},\ }\bibfield
  {title} {\bibinfo {title} {A locally adaptive phase aberration correction
  {(LAPAC)} method for synthetic aperture sequences},\ }\href@noop {}
  {\bibfield  {journal} {\bibinfo  {journal} {Ultrason. Imag.}\ }\textbf
  {\bibinfo {volume} {41}},\ \bibinfo {pages} {3} (\bibinfo {year}
  {2019})}\BibitemShut {NoStop}%
\bibitem [{\citenamefont {Jaeger}\ \emph
  {et~al.}(2015{\natexlab{a}})\citenamefont {Jaeger}, \citenamefont {Robinson},
  \citenamefont {G\"{u}nhan~Akar\c{c}ay},\ and\ \citenamefont
  {Frenz}}]{Jaeger2015SosMap}%
  \BibitemOpen
  \bibfield  {author} {\bibinfo {author} {\bibfnamefont {M.}~\bibnamefont
  {Jaeger}}, \bibinfo {author} {\bibfnamefont {E.}~\bibnamefont {Robinson}},
  \bibinfo {author} {\bibfnamefont {H.}~\bibnamefont
  {G\"{u}nhan~Akar\c{c}ay}},\ and\ \bibinfo {author} {\bibfnamefont
  {M.}~\bibnamefont {Frenz}},\ }\bibfield  {title} {\bibinfo {title} {Full
  correction for spatially distributed speed-of-sound in echo ultrasound based
  on measuring aberration delays via transmit beam steering},\ }\href@noop {}
  {\bibfield  {journal} {\bibinfo  {journal} {Phys. Med. Biol.}\ }\textbf
  {\bibinfo {volume} {60}},\ \bibinfo {pages} {4497} (\bibinfo {year}
  {2015}{\natexlab{a}})}\BibitemShut {NoStop}%
\bibitem [{\citenamefont {Durduran}\ \emph {et~al.}(2010)\citenamefont
  {Durduran}, \citenamefont {Choe}, \citenamefont {Baker},\ and\ \citenamefont
  {Yodh}}]{Durduran2010}%
  \BibitemOpen
  \bibfield  {author} {\bibinfo {author} {\bibfnamefont {T.}~\bibnamefont
  {Durduran}}, \bibinfo {author} {\bibfnamefont {R.}~\bibnamefont {Choe}},
  \bibinfo {author} {\bibfnamefont {W.~B.}\ \bibnamefont {Baker}},\ and\
  \bibinfo {author} {\bibfnamefont {A.~G.}\ \bibnamefont {Yodh}},\ }\bibfield
  {title} {\bibinfo {title} {Diffuse optics for tissue monitoring and
  tomography.},\ }\href@noop {} {\bibfield  {journal} {\bibinfo  {journal}
  {Rep. Prog. Phys.}\ }\textbf {\bibinfo {volume} {73}},\ \bibinfo {pages}
  {076701} (\bibinfo {year} {2010})}\BibitemShut {NoStop}%
\bibitem [{\citenamefont {Montaldo}\ \emph {et~al.}(2009)\citenamefont
  {Montaldo}, \citenamefont {Tanter}, \citenamefont {Bercoff}, \citenamefont
  {Benech},\ and\ \citenamefont {Fink}}]{Montaldo2009}%
  \BibitemOpen
  \bibfield  {author} {\bibinfo {author} {\bibfnamefont {G.}~\bibnamefont
  {Montaldo}}, \bibinfo {author} {\bibfnamefont {M.}~\bibnamefont {Tanter}},
  \bibinfo {author} {\bibfnamefont {J.}~\bibnamefont {Bercoff}}, \bibinfo
  {author} {\bibfnamefont {N.}~\bibnamefont {Benech}},\ and\ \bibinfo {author}
  {\bibfnamefont {M.}~\bibnamefont {Fink}},\ }\bibfield  {title} {\bibinfo
  {title} {{Coherent plane wave compounding for very high frame rate
  ultrasonography and transient elastography}},\ }\href@noop {} {\bibfield
  {journal} {\bibinfo  {journal} {IEEE Trans. Ultrason. Ferroelectr. Freq.
  Control}\ }\textbf {\bibinfo {volume} {56}},\ \bibinfo {pages} {489}
  (\bibinfo {year} {2009})}\BibitemShut {NoStop}%
\bibitem [{\citenamefont {Mosk}\ and\ \citenamefont {van
  Putten}(2010)}]{vanPutten2010}%
  \BibitemOpen
  \bibfield  {author} {\bibinfo {author} {\bibfnamefont {A.~P.}\ \bibnamefont
  {Mosk}}\ and\ \bibinfo {author} {\bibfnamefont {E.~G.}\ \bibnamefont {van
  Putten}},\ }\bibfield  {title} {\bibinfo {title} {The information age in
  optics: Measuring the transmission matrix},\ }\href@noop {} {\bibfield
  {journal} {\bibinfo  {journal} {Physics}\ }\textbf {\bibinfo {volume} {3}},\
  \bibinfo {pages} {22} (\bibinfo {year} {2010})}\BibitemShut {NoStop}%
\bibitem [{\citenamefont {Provost}\ \emph {et~al.}(2014)\citenamefont
  {Provost}, \citenamefont {Papadacci}, \citenamefont {Arango}, \citenamefont
  {Imbault}, \citenamefont {Fink}, \citenamefont {Gennisson}, \citenamefont
  {Tanter},\ and\ \citenamefont {Pernot}}]{Provost2014}%
  \BibitemOpen
  \bibfield  {author} {\bibinfo {author} {\bibfnamefont {J.}~\bibnamefont
  {Provost}}, \bibinfo {author} {\bibfnamefont {C.}~\bibnamefont {Papadacci}},
  \bibinfo {author} {\bibfnamefont {J.~E.}\ \bibnamefont {Arango}}, \bibinfo
  {author} {\bibfnamefont {M.}~\bibnamefont {Imbault}}, \bibinfo {author}
  {\bibfnamefont {M.}~\bibnamefont {Fink}}, \bibinfo {author} {\bibfnamefont
  {J.-L.}\ \bibnamefont {Gennisson}}, \bibinfo {author} {\bibfnamefont
  {M.}~\bibnamefont {Tanter}},\ and\ \bibinfo {author} {\bibfnamefont
  {M.}~\bibnamefont {Pernot}},\ }\bibfield  {title} {\bibinfo {title} {{3D}
  ultrafast ultrasound imaging in vivo},\ }\href@noop {} {\bibfield  {journal}
  {\bibinfo  {journal} {Phys. Med. Biol.}\ }\textbf {\bibinfo {volume} {59}},\
  \bibinfo {pages} {L1} (\bibinfo {year} {2014})}\BibitemShut {NoStop}%
\bibitem [{\citenamefont {Varslot}\ \emph {et~al.}(2004)\citenamefont
  {Varslot}, \citenamefont {Krogstad}, \citenamefont {Mo},\ and\ \citenamefont
  {Angelsen}}]{Varslot2004}%
  \BibitemOpen
  \bibfield  {author} {\bibinfo {author} {\bibfnamefont {T.}~\bibnamefont
  {Varslot}}, \bibinfo {author} {\bibfnamefont {H.}~\bibnamefont {Krogstad}},
  \bibinfo {author} {\bibfnamefont {E.}~\bibnamefont {Mo}},\ and\ \bibinfo
  {author} {\bibfnamefont {B.~A.}\ \bibnamefont {Angelsen}},\ }\bibfield
  {title} {\bibinfo {title} {Eigenfunction analysis of stochastic backscatter
  for characterization of acoustic aberration in medical ultrasound imaging},\
  }\href@noop {} {\bibfield  {journal} {\bibinfo  {journal} {J. Acoust. Soc.
  Am.}\ }\textbf {\bibinfo {volume} {115}},\ \bibinfo {pages} {3068} (\bibinfo
  {year} {2004})}\BibitemShut {NoStop}%
\bibitem [{\citenamefont {Robert}\ and\ \citenamefont
  {Fink}(2008)}]{Robert2008}%
  \BibitemOpen
  \bibfield  {author} {\bibinfo {author} {\bibfnamefont {J.-L.}\ \bibnamefont
  {Robert}}\ and\ \bibinfo {author} {\bibfnamefont {M.}~\bibnamefont {Fink}},\
  }\bibfield  {title} {\bibinfo {title} {Green's function estimation in speckle
  using the decomposition of the time reversal operator: Application to
  aberration correction in medical imaging},\ }\href@noop {} {\bibfield
  {journal} {\bibinfo  {journal} {J. Acoust. Soc. Am.}\ }\textbf {\bibinfo
  {volume} {123}},\ \bibinfo {pages} {866} (\bibinfo {year}
  {2008})}\BibitemShut {NoStop}%
\bibitem [{\citenamefont {Kang}\ \emph {et~al.}(2017)\citenamefont {Kang},
  \citenamefont {Kang}, \citenamefont {Jeong}, \citenamefont {Kwon},
  \citenamefont {Yang}, \citenamefont {Hong}, \citenamefont {Kim},
  \citenamefont {Song}, \citenamefont {Park}, \citenamefont {Lee},
  \citenamefont {Kim}, \citenamefont {Kim},\ and\ \citenamefont
  {Choi}}]{Kang2017}%
  \BibitemOpen
  \bibfield  {author} {\bibinfo {author} {\bibfnamefont {S.}~\bibnamefont
  {Kang}}, \bibinfo {author} {\bibfnamefont {P.}~\bibnamefont {Kang}}, \bibinfo
  {author} {\bibfnamefont {S.}~\bibnamefont {Jeong}}, \bibinfo {author}
  {\bibfnamefont {Y.}~\bibnamefont {Kwon}}, \bibinfo {author} {\bibfnamefont
  {T.~D.}\ \bibnamefont {Yang}}, \bibinfo {author} {\bibfnamefont {J.~H.}\
  \bibnamefont {Hong}}, \bibinfo {author} {\bibfnamefont {M.}~\bibnamefont
  {Kim}}, \bibinfo {author} {\bibfnamefont {K.-D.}\ \bibnamefont {Song}},
  \bibinfo {author} {\bibfnamefont {J.~H.}\ \bibnamefont {Park}}, \bibinfo
  {author} {\bibfnamefont {J.~H.}\ \bibnamefont {Lee}}, \bibinfo {author}
  {\bibfnamefont {M.~J.}\ \bibnamefont {Kim}}, \bibinfo {author} {\bibfnamefont
  {K.~H.}\ \bibnamefont {Kim}},\ and\ \bibinfo {author} {\bibfnamefont
  {W.}~\bibnamefont {Choi}},\ }\bibfield  {title} {\bibinfo {title}
  {{High-resolution adaptive optical imaging within thick scattering media
  using closed-loop accumulation of single scattering}},\ }\href@noop {}
  {\bibfield  {journal} {\bibinfo  {journal} {Nat. Commun.}\ }\textbf {\bibinfo
  {volume} {8}},\ \bibinfo {pages} {2157} (\bibinfo {year} {2017})}\BibitemShut
  {NoStop}%
\bibitem [{\citenamefont {Badon}\ \emph {et~al.}(2019)\citenamefont {Badon},
  \citenamefont {Barolle}, \citenamefont {Irsch}, \citenamefont {Boccara},
  \citenamefont {Fink},\ and\ \citenamefont {Aubry}}]{Badon2019}%
  \BibitemOpen
  \bibfield  {author} {\bibinfo {author} {\bibfnamefont {A.}~\bibnamefont
  {Badon}}, \bibinfo {author} {\bibfnamefont {V.}~\bibnamefont {Barolle}},
  \bibinfo {author} {\bibfnamefont {K.}~\bibnamefont {Irsch}}, \bibinfo
  {author} {\bibfnamefont {A.~C.}\ \bibnamefont {Boccara}}, \bibinfo {author}
  {\bibfnamefont {M.}~\bibnamefont {Fink}},\ and\ \bibinfo {author}
  {\bibfnamefont {A.}~\bibnamefont {Aubry}},\ }\bibfield  {title} {\bibinfo
  {title} {{Distortion matrix concept for deep imaging in optical coherence
  microscopy}},\ }\href@noop {} {\bibfield  {journal} {\bibinfo  {journal}
  {arXiv: 1910.07252}\ } (\bibinfo {year} {2019})}\BibitemShut {NoStop}%
\bibitem [{\citenamefont {Aubry}\ and\ \citenamefont
  {Derode}(2009)}]{Aubry2009a}%
  \BibitemOpen
  \bibfield  {author} {\bibinfo {author} {\bibfnamefont {A.}~\bibnamefont
  {Aubry}}\ and\ \bibinfo {author} {\bibfnamefont {A.}~\bibnamefont {Derode}},\
  }\bibfield  {title} {\bibinfo {title} {{Random matrix theory applied to
  acoustic backscattering and Imaging In complex media}},\ }\href@noop {}
  {\bibfield  {journal} {\bibinfo  {journal} {Phys. Rev. Lett.}\ }\textbf
  {\bibinfo {volume} {102}},\ \bibinfo {pages} {084301} (\bibinfo {year}
  {2009})}\BibitemShut {NoStop}%
\bibitem [{\citenamefont {Aubry}\ and\ \citenamefont
  {Derode}(2011)}]{Aubry2011}%
  \BibitemOpen
  \bibfield  {author} {\bibinfo {author} {\bibfnamefont {A.}~\bibnamefont
  {Aubry}}\ and\ \bibinfo {author} {\bibfnamefont {A.}~\bibnamefont {Derode}},\
  }\bibfield  {title} {\bibinfo {title} {{Multiple scattering of ultrasound in
  weakly inhomogeneous media: Application to human soft tissues}},\ }\href@noop
  {} {\bibfield  {journal} {\bibinfo  {journal} {J. Acoust. Soc. Am.}\ }\textbf
  {\bibinfo {volume} {129}},\ \bibinfo {pages} {225} (\bibinfo {year}
  {2011})}\BibitemShut {NoStop}%
\bibitem [{\citenamefont {Kang}\ \emph {et~al.}(2015)\citenamefont {Kang},
  \citenamefont {Jeong}, \citenamefont {Choi}, \citenamefont {Yang},
  \citenamefont {Joo}, \citenamefont {Lee}, \citenamefont {Lim}, \citenamefont
  {Park},\ and\ \citenamefont {Choi}}]{Kang2015}%
  \BibitemOpen
  \bibfield  {author} {\bibinfo {author} {\bibfnamefont {S.}~\bibnamefont
  {Kang}}, \bibinfo {author} {\bibfnamefont {S.}~\bibnamefont {Jeong}},
  \bibinfo {author} {\bibfnamefont {H.}~\bibnamefont {Choi}, \bibfnamefont
  {W.~Ko}}, \bibinfo {author} {\bibfnamefont {T.~D.}\ \bibnamefont {Yang}},
  \bibinfo {author} {\bibfnamefont {J.~H.}\ \bibnamefont {Joo}}, \bibinfo
  {author} {\bibfnamefont {J.-S.}\ \bibnamefont {Lee}}, \bibinfo {author}
  {\bibfnamefont {Y.-S.}\ \bibnamefont {Lim}}, \bibinfo {author} {\bibfnamefont
  {Q.-H.}\ \bibnamefont {Park}},\ and\ \bibinfo {author} {\bibfnamefont
  {W.}~\bibnamefont {Choi}},\ }\bibfield  {title} {\bibinfo {title} {{Imaging
  deep within a scattering medium using collective accumulation of
  single-scattered waves}},\ }\href@noop {} {\bibfield  {journal} {\bibinfo
  {journal} {Nat. Photonics}\ }\textbf {\bibinfo {volume} {9}},\ \bibinfo
  {pages} {253} (\bibinfo {year} {2015})}\BibitemShut {NoStop}%
\bibitem [{\citenamefont {Badon}\ \emph
  {et~al.}(2016{\natexlab{a}})\citenamefont {Badon}, \citenamefont {Li},
  \citenamefont {Lerosey}, \citenamefont {Boccara}, \citenamefont {Fink},\ and\
  \citenamefont {Aubry}}]{Badon2016}%
  \BibitemOpen
  \bibfield  {author} {\bibinfo {author} {\bibfnamefont {A.}~\bibnamefont
  {Badon}}, \bibinfo {author} {\bibfnamefont {D.}~\bibnamefont {Li}}, \bibinfo
  {author} {\bibfnamefont {G.}~\bibnamefont {Lerosey}}, \bibinfo {author}
  {\bibfnamefont {A.~C.}\ \bibnamefont {Boccara}}, \bibinfo {author}
  {\bibfnamefont {M.}~\bibnamefont {Fink}},\ and\ \bibinfo {author}
  {\bibfnamefont {A.}~\bibnamefont {Aubry}},\ }\bibfield  {title} {\bibinfo
  {title} {{Smart optical coherence tomography for ultra-deep imaging through
  highly scattering media}},\ }\href@noop {} {\bibfield  {journal} {\bibinfo
  {journal} {Sci. Adv.}\ }\textbf {\bibinfo {volume} {2}},\ \bibinfo {pages}
  {e1600370} (\bibinfo {year} {2016}{\natexlab{a}})}\BibitemShut {NoStop}%
\bibitem [{\citenamefont {Blondel}\ \emph {et~al.}(2018)\citenamefont
  {Blondel}, \citenamefont {Chaput}, \citenamefont {Derode}, \citenamefont
  {Campillo},\ and\ \citenamefont {Aubry}}]{Blondel2018}%
  \BibitemOpen
  \bibfield  {author} {\bibinfo {author} {\bibfnamefont {T.}~\bibnamefont
  {Blondel}}, \bibinfo {author} {\bibfnamefont {J.}~\bibnamefont {Chaput}},
  \bibinfo {author} {\bibfnamefont {A.}~\bibnamefont {Derode}}, \bibinfo
  {author} {\bibfnamefont {M.}~\bibnamefont {Campillo}},\ and\ \bibinfo
  {author} {\bibfnamefont {A.}~\bibnamefont {Aubry}},\ }\bibfield  {title}
  {\bibinfo {title} {{Matrix approach of seismic imaging: Application to the
  Erebus volcano, Antarctica}},\ }\href@noop {} {\bibfield  {journal} {\bibinfo
   {journal} {J. Geophys. Res.: Solid Earth}\ }\textbf {\bibinfo {volume}
  {123}},\ \bibinfo {pages} {10936} (\bibinfo {year} {2018})}\BibitemShut
  {NoStop}%
\bibitem [{\citenamefont {Jaeger}\ \emph
  {et~al.}(2015{\natexlab{b}})\citenamefont {Jaeger}, \citenamefont {Held},
  \citenamefont {Peeters}, \citenamefont {Preisser}, \citenamefont
  {Gr{\"{u}}nig},\ and\ \citenamefont {Frenz}}]{Jaeger2015}%
  \BibitemOpen
  \bibfield  {author} {\bibinfo {author} {\bibfnamefont {M.}~\bibnamefont
  {Jaeger}}, \bibinfo {author} {\bibfnamefont {G.}~\bibnamefont {Held}},
  \bibinfo {author} {\bibfnamefont {S.}~\bibnamefont {Peeters}}, \bibinfo
  {author} {\bibfnamefont {S.}~\bibnamefont {Preisser}}, \bibinfo {author}
  {\bibfnamefont {M.}~\bibnamefont {Gr{\"{u}}nig}},\ and\ \bibinfo {author}
  {\bibfnamefont {M.}~\bibnamefont {Frenz}},\ }\bibfield  {title} {\bibinfo
  {title} {{Computed ultrasound tomography in echo mode for imaging speed of
  sound using pulse-echo sonography: Proof of principle}},\ }\href@noop {}
  {\bibfield  {journal} {\bibinfo  {journal} {Ultrasound Med. Biol.}\ }\textbf
  {\bibinfo {volume} {41}},\ \bibinfo {pages} {235} (\bibinfo {year}
  {2015}{\natexlab{b}})}\BibitemShut {NoStop}%
\bibitem [{\citenamefont {Imbault}\ \emph {et~al.}(2017)\citenamefont
  {Imbault}, \citenamefont {Faccinetto}, \citenamefont {Osmanski},
  \citenamefont {Tissier}, \citenamefont {Deffieux}, \citenamefont {Gennisson},
  \citenamefont {Vilgrain},\ and\ \citenamefont {Tanter}}]{Imbault2017}%
  \BibitemOpen
  \bibfield  {author} {\bibinfo {author} {\bibfnamefont {M.}~\bibnamefont
  {Imbault}}, \bibinfo {author} {\bibfnamefont {A.}~\bibnamefont {Faccinetto}},
  \bibinfo {author} {\bibfnamefont {B.-F.}\ \bibnamefont {Osmanski}}, \bibinfo
  {author} {\bibfnamefont {A.}~\bibnamefont {Tissier}}, \bibinfo {author}
  {\bibfnamefont {T.}~\bibnamefont {Deffieux}}, \bibinfo {author}
  {\bibfnamefont {J.-L.}\ \bibnamefont {Gennisson}}, \bibinfo {author}
  {\bibfnamefont {V.}~\bibnamefont {Vilgrain}},\ and\ \bibinfo {author}
  {\bibfnamefont {M.}~\bibnamefont {Tanter}},\ }\bibfield  {title} {\bibinfo
  {title} {{Robust sound speed estimation for ultrasound-based hepatic
  steatosis assessment}},\ }\href@noop {} {\bibfield  {journal} {\bibinfo
  {journal} {Phys. Med. Biol.}\ }\textbf {\bibinfo {volume} {62}},\ \bibinfo
  {pages} {3582} (\bibinfo {year} {2017})}\BibitemShut {NoStop}%
\bibitem [{\citenamefont {St\"ahli}\ \emph {et~al.}(2019)\citenamefont
  {St\"ahli}, \citenamefont {Kuriakose}, \citenamefont {Frenz},\ and\
  \citenamefont {Jaeger}}]{Stahli2019}%
  \BibitemOpen
  \bibfield  {author} {\bibinfo {author} {\bibfnamefont {P.}~\bibnamefont
  {St\"ahli}}, \bibinfo {author} {\bibfnamefont {M.}~\bibnamefont {Kuriakose}},
  \bibinfo {author} {\bibfnamefont {M.}~\bibnamefont {Frenz}},\ and\ \bibinfo
  {author} {\bibfnamefont {M.}~\bibnamefont {Jaeger}},\ }\bibfield  {title}
  {\bibinfo {title} {Forward model for quantitative pulse-echo speed-of-sound
  imaging},\ }\href@noop {} {\bibfield  {journal} {\bibinfo  {journal} {arXiv:
  1902.10639}\ } (\bibinfo {year} {2019})}\BibitemShut {NoStop}%
\bibitem [{\citenamefont {Aubry}\ \emph {et~al.}(2008)\citenamefont {Aubry},
  \citenamefont {Derode},\ and\ \citenamefont {Padilla}}]{Aubry2008}%
  \BibitemOpen
  \bibfield  {author} {\bibinfo {author} {\bibfnamefont {A.}~\bibnamefont
  {Aubry}}, \bibinfo {author} {\bibfnamefont {A.}~\bibnamefont {Derode}},\ and\
  \bibinfo {author} {\bibfnamefont {F.}~\bibnamefont {Padilla}},\ }\bibfield
  {title} {\bibinfo {title} {{Local measurements of the diffusion constant in
  multiple scattering media: Application to human trabecular bone imaging}},\
  }\href
  {http://scitation.aip.org/content/aip/journal/apl/92/12/10.1063/1.2901379}
  {\bibfield  {journal} {\bibinfo  {journal} {Appl. Phys. Lett.}\ }\textbf
  {\bibinfo {volume} {92}},\ \bibinfo {pages} {124101} (\bibinfo {year}
  {2008})}\BibitemShut {NoStop}%
\bibitem [{\citenamefont {Mohanty}\ \emph {et~al.}(2017)\citenamefont
  {Mohanty}, \citenamefont {Blackwell}, \citenamefont {Egan},\ and\
  \citenamefont {Muller}}]{Mohanty2017}%
  \BibitemOpen
  \bibfield  {author} {\bibinfo {author} {\bibfnamefont {K.}~\bibnamefont
  {Mohanty}}, \bibinfo {author} {\bibfnamefont {J.}~\bibnamefont {Blackwell}},
  \bibinfo {author} {\bibfnamefont {T.}~\bibnamefont {Egan}},\ and\ \bibinfo
  {author} {\bibfnamefont {M.}~\bibnamefont {Muller}},\ }\bibfield  {title}
  {\bibinfo {title} {Characterization of the lung parenchyma using ultrasound
  multiple scattering},\ }\href@noop {} {\bibfield  {journal} {\bibinfo
  {journal} {Ultrasound Med. Biol.}\ }\textbf {\bibinfo {volume} {43}},\
  \bibinfo {pages} {993} (\bibinfo {year} {2017})}\BibitemShut {NoStop}%
\bibitem [{\citenamefont {Suzuki}\ \emph {et~al.}(1992)\citenamefont {Suzuki},
  \citenamefont {Hayashi}, \citenamefont {Sasaki}, \citenamefont {Kono},
  \citenamefont {Kasahara}, \citenamefont {Fusamoto}, \citenamefont {Imai},\
  and\ \citenamefont {Kamada}}]{Suzuki1992}%
  \BibitemOpen
  \bibfield  {author} {\bibinfo {author} {\bibfnamefont {K.}~\bibnamefont
  {Suzuki}}, \bibinfo {author} {\bibfnamefont {N.}~\bibnamefont {Hayashi}},
  \bibinfo {author} {\bibfnamefont {Y.}~\bibnamefont {Sasaki}}, \bibinfo
  {author} {\bibfnamefont {M.}~\bibnamefont {Kono}}, \bibinfo {author}
  {\bibfnamefont {A.}~\bibnamefont {Kasahara}}, \bibinfo {author}
  {\bibfnamefont {H.}~\bibnamefont {Fusamoto}}, \bibinfo {author}
  {\bibfnamefont {Y.}~\bibnamefont {Imai}},\ and\ \bibinfo {author}
  {\bibfnamefont {T.}~\bibnamefont {Kamada}},\ }\bibfield  {title} {\bibinfo
  {title} {Dependence of ultrasonic attenuation of liver on pathologic fat and
  fibrosis: examination with experimental fatty liver and liver fibrosis
  models},\ }\href@noop {} {\bibfield  {journal} {\bibinfo  {journal}
  {Ultrasound Med. Biol.}\ }\textbf {\bibinfo {volume} {18}},\ \bibinfo {pages}
  {657} (\bibinfo {year} {1992})}\BibitemShut {NoStop}%
\bibitem [{\citenamefont {Sasso}\ \emph {et~al.}(2010)\citenamefont {Sasso},
  \citenamefont {Beaugrand}, \citenamefont {de~Ledinghen}, \citenamefont
  {Douvin}, \citenamefont {Marcellin}, \citenamefont {Poupon}, \citenamefont
  {Sandrin},\ and\ \citenamefont {Miette}}]{Sasso2010}%
  \BibitemOpen
  \bibfield  {author} {\bibinfo {author} {\bibfnamefont {M.}~\bibnamefont
  {Sasso}}, \bibinfo {author} {\bibfnamefont {M.}~\bibnamefont {Beaugrand}},
  \bibinfo {author} {\bibfnamefont {V.}~\bibnamefont {de~Ledinghen}}, \bibinfo
  {author} {\bibfnamefont {C.}~\bibnamefont {Douvin}}, \bibinfo {author}
  {\bibfnamefont {P.}~\bibnamefont {Marcellin}}, \bibinfo {author}
  {\bibfnamefont {R.}~\bibnamefont {Poupon}}, \bibinfo {author} {\bibfnamefont
  {L.}~\bibnamefont {Sandrin}},\ and\ \bibinfo {author} {\bibfnamefont
  {V.}~\bibnamefont {Miette}},\ }\bibfield  {title} {\bibinfo {title}
  {{Controlled attenuation parameter (CAP): a novel VCTE$^{\mbox{TM}}$ guided
  ultrasonic attenuation measurement for the evaluation of hepatic steatosis:
  preliminary study and validation in a cohort of patients with chronic liver
  disease from various causes}},\ }\href@noop {} {\bibfield  {journal}
  {\bibinfo  {journal} {Ultrasound Med. Biol.}\ }\textbf {\bibinfo {volume}
  {36}},\ \bibinfo {pages} {1825} (\bibinfo {year} {2010})}\BibitemShut
  {NoStop}%
\bibitem [{\citenamefont {Bamber}\ and\ \citenamefont
  {Hill}(1981)}]{Bamber1981}%
  \BibitemOpen
  \bibfield  {author} {\bibinfo {author} {\bibfnamefont {J.~C.}\ \bibnamefont
  {Bamber}}\ and\ \bibinfo {author} {\bibfnamefont {C.~R.}\ \bibnamefont
  {Hill}},\ }\bibfield  {title} {\bibinfo {title} {{Acoustic properties of
  normal and cancerous human liver- I. Dependence on pathological condition}},\
  }\href@noop {} {\bibfield  {journal} {\bibinfo  {journal} {Ultrasound Med.
  Biol.}\ }\textbf {\bibinfo {volume} {7}},\ \bibinfo {pages} {121} (\bibinfo
  {year} {1981})}\BibitemShut {NoStop}%
\bibitem [{\citenamefont {Bamber}\ \emph {et~al.}(1981)\citenamefont {Bamber},
  \citenamefont {Hill},\ and\ \citenamefont {King}}]{Bamber1981b}%
  \BibitemOpen
  \bibfield  {author} {\bibinfo {author} {\bibfnamefont {J.~C.}\ \bibnamefont
  {Bamber}}, \bibinfo {author} {\bibfnamefont {C.~R.}\ \bibnamefont {Hill}},\
  and\ \bibinfo {author} {\bibfnamefont {J.~A.}\ \bibnamefont {King}},\
  }\bibfield  {title} {\bibinfo {title} {{Acoustic properties of normal and
  cancerous human liver-II. Dependence on tissue structure}},\ }\href@noop {}
  {\bibfield  {journal} {\bibinfo  {journal} {Ultrasound Med. Biol.}\ }\textbf
  {\bibinfo {volume} {7}},\ \bibinfo {pages} {135} (\bibinfo {year}
  {1981})}\BibitemShut {NoStop}%
\bibitem [{\citenamefont {Chen}\ \emph {et~al.}(1987)\citenamefont {Chen},
  \citenamefont {Robinson}, \citenamefont {Wilson}, \citenamefont {Griffiths},
  \citenamefont {Manoharan},\ and\ \citenamefont {Doust}}]{Chen1987}%
  \BibitemOpen
  \bibfield  {author} {\bibinfo {author} {\bibfnamefont {C.~F.}\ \bibnamefont
  {Chen}}, \bibinfo {author} {\bibfnamefont {D.~E.}\ \bibnamefont {Robinson}},
  \bibinfo {author} {\bibfnamefont {L.~S.}\ \bibnamefont {Wilson}}, \bibinfo
  {author} {\bibfnamefont {K.~A.}\ \bibnamefont {Griffiths}}, \bibinfo {author}
  {\bibfnamefont {A.}~\bibnamefont {Manoharan}},\ and\ \bibinfo {author}
  {\bibfnamefont {B.~D.}\ \bibnamefont {Doust}},\ }\bibfield  {title} {\bibinfo
  {title} {Clinical sound speed measurement in liver and spleen in vivo},\
  }\href@noop {} {\bibfield  {journal} {\bibinfo  {journal} {Ultrason.
  Imaging}\ }\textbf {\bibinfo {volume} {9}},\ \bibinfo {pages} {221} (\bibinfo
  {year} {1987})}\BibitemShut {NoStop}%
\bibitem [{\citenamefont {Duck}(1990)}]{Duck1990}%
  \BibitemOpen
  \bibfield  {author} {\bibinfo {author} {\bibfnamefont {F.~A.}\ \bibnamefont
  {Duck}},\ }\href@noop {} {\emph {\bibinfo {title} {Physical Properties of
  Tissues}}}\ (\bibinfo  {publisher} {Elsevier, Amsterdam},\ \bibinfo {year}
  {1990})\ pp.\ \bibinfo {pages} {73 -- 135}\BibitemShut {NoStop}%
\bibitem [{\citenamefont {Schurr}\ \emph {et~al.}(2011)\citenamefont {Schurr},
  \citenamefont {Sabra},\ and\ \citenamefont {Jacobs}}]{Schurr2011}%
  \BibitemOpen
  \bibfield  {author} {\bibinfo {author} {\bibfnamefont {J.~Y.}\ \bibnamefont
  {Schurr}, \bibfnamefont {D.~P.~Kim}}, \bibinfo {author} {\bibfnamefont
  {K.~G.}\ \bibnamefont {Sabra}},\ and\ \bibinfo {author} {\bibfnamefont
  {L.~J.}\ \bibnamefont {Jacobs}},\ }\bibfield  {title} {\bibinfo {title}
  {Damage detection in concrete using coda wave interferometry},\ }\href@noop
  {} {\bibfield  {journal} {\bibinfo  {journal} {NDT{\&}E Int.}\ }\textbf
  {\bibinfo {volume} {44}},\ \bibinfo {pages} {728} (\bibinfo {year}
  {2011})}\BibitemShut {NoStop}%
\bibitem [{\citenamefont {Shajahan}\ \emph {et~al.}(2014)\citenamefont
  {Shajahan}, \citenamefont {Rupin}, \citenamefont {Aubry}, \citenamefont
  {Chassignole}, \citenamefont {Fouquet},\ and\ \citenamefont
  {Derode}}]{Shajahan2014b}%
  \BibitemOpen
  \bibfield  {author} {\bibinfo {author} {\bibfnamefont {S.}~\bibnamefont
  {Shajahan}}, \bibinfo {author} {\bibfnamefont {F.}~\bibnamefont {Rupin}},
  \bibinfo {author} {\bibfnamefont {A.}~\bibnamefont {Aubry}}, \bibinfo
  {author} {\bibfnamefont {B.}~\bibnamefont {Chassignole}}, \bibinfo {author}
  {\bibfnamefont {T.}~\bibnamefont {Fouquet}},\ and\ \bibinfo {author}
  {\bibfnamefont {A.}~\bibnamefont {Derode}},\ }\bibfield  {title} {\bibinfo
  {title} {Comparison between experimental and 2-d numerical studies of
  multiple scattering in inconel600® by means of array probes},\ }\href@noop
  {} {\bibfield  {journal} {\bibinfo  {journal} {Ultrasonics}\ }\textbf
  {\bibinfo {volume} {54}},\ \bibinfo {pages} {358} (\bibinfo {year}
  {2014})}\BibitemShut {NoStop}%
\bibitem [{\citenamefont {Zhang}\ \emph {et~al.}(2016)\citenamefont {Zhang},
  \citenamefont {Planes}, \citenamefont {Larose},\ and\ \citenamefont
  {Obermann}}]{Zhang2016}%
  \BibitemOpen
  \bibfield  {author} {\bibinfo {author} {\bibfnamefont {Y.}~\bibnamefont
  {Zhang}}, \bibinfo {author} {\bibfnamefont {T.}~\bibnamefont {Planes}},
  \bibinfo {author} {\bibfnamefont {E.}~\bibnamefont {Larose}},\ and\ \bibinfo
  {author} {\bibfnamefont {A.}~\bibnamefont {Obermann}},\ }\bibfield  {title}
  {\bibinfo {title} {Diffuse ultrasound monitoring of stress and damage
  development on a 15-ton concrete beam},\ }\href@noop {} {\bibfield  {journal}
  {\bibinfo  {journal} {J. Acoust. Soc. Am.}\ }\textbf {\bibinfo {volume}
  {139}},\ \bibinfo {pages} {1691} (\bibinfo {year} {2016})}\BibitemShut
  {NoStop}%
\bibitem [{\citenamefont {Sato}\ \emph {et~al.}(2012)\citenamefont {Sato},
  \citenamefont {Fehler},\ and\ \citenamefont {Maeda}}]{Sato2012}%
  \BibitemOpen
  \bibfield  {author} {\bibinfo {author} {\bibfnamefont {H.}~\bibnamefont
  {Sato}}, \bibinfo {author} {\bibfnamefont {M.~C.}\ \bibnamefont {Fehler}},\
  and\ \bibinfo {author} {\bibfnamefont {T.}~\bibnamefont {Maeda}},\
  }\href@noop {} {\emph {\bibinfo {title} {Seismic Wave Propagation and
  Scattering in the Heterogeneous Earth : Second Edition}}}\ (\bibinfo
  {publisher} {Springer-Verlag, Berlin},\ \bibinfo {year} {2012})\BibitemShut
  {NoStop}%
\bibitem [{\citenamefont {Chaput}\ \emph {et~al.}(2015)\citenamefont {Chaput},
  \citenamefont {Campillo}, \citenamefont {Aster}, \citenamefont {Roux},
  \citenamefont {Kyle}, \citenamefont {Knox},\ and\ \citenamefont
  {Czoski}}]{Chaput2015}%
  \BibitemOpen
  \bibfield  {author} {\bibinfo {author} {\bibfnamefont {J.}~\bibnamefont
  {Chaput}}, \bibinfo {author} {\bibfnamefont {M.}~\bibnamefont {Campillo}},
  \bibinfo {author} {\bibfnamefont {R.~C.}\ \bibnamefont {Aster}}, \bibinfo
  {author} {\bibfnamefont {P.}~\bibnamefont {Roux}}, \bibinfo {author}
  {\bibfnamefont {P.~R.}\ \bibnamefont {Kyle}}, \bibinfo {author}
  {\bibfnamefont {H.}~\bibnamefont {Knox}},\ and\ \bibinfo {author}
  {\bibfnamefont {P.}~\bibnamefont {Czoski}},\ }\bibfield  {title} {\bibinfo
  {title} {{Multiple scattering from icequakes at Erebus volcano, Antarctica:
  Implications for imaging at glaciated volcanoes}},\ }\href@noop {} {\bibfield
   {journal} {\bibinfo  {journal} {J. Geophys. Res.}\ }\textbf {\bibinfo
  {volume} {120}},\ \bibinfo {pages} {1129} (\bibinfo {year}
  {2015})}\BibitemShut {NoStop}%
\bibitem [{\citenamefont {Mayor}\ \emph {et~al.}(2018)\citenamefont {Mayor},
  \citenamefont {Traversa}, \citenamefont {Calvet},\ and\ \citenamefont
  {Margerin}}]{Mayor2018}%
  \BibitemOpen
  \bibfield  {author} {\bibinfo {author} {\bibfnamefont {J.}~\bibnamefont
  {Mayor}}, \bibinfo {author} {\bibfnamefont {P.}~\bibnamefont {Traversa}},
  \bibinfo {author} {\bibfnamefont {M.}~\bibnamefont {Calvet}},\ and\ \bibinfo
  {author} {\bibfnamefont {L.}~\bibnamefont {Margerin}},\ }\bibfield  {title}
  {\bibinfo {title} {Tomography of crustal seismic attenuation in metropolitan
  france: Implications for seismicity analysis},\ }\href@noop {} {\bibfield
  {journal} {\bibinfo  {journal} {Bull. Earthquake Eng.}\ }\textbf {\bibinfo
  {volume} {16}} (\bibinfo {year} {2018})}\BibitemShut {NoStop}%
\bibitem [{\citenamefont {Prada}\ and\ \citenamefont {Fink}(1994)}]{Prada1994}%
  \BibitemOpen
  \bibfield  {author} {\bibinfo {author} {\bibfnamefont {C.}~\bibnamefont
  {Prada}}\ and\ \bibinfo {author} {\bibfnamefont {M.}~\bibnamefont {Fink}},\
  }\bibfield  {title} {\bibinfo {title} {{Eigenmodes of the time reversal
  operator: A solution to selective focusing in multiple-target media}},\
  }\href@noop {} {\bibfield  {journal} {\bibinfo  {journal} {Wave Motion}\
  }\textbf {\bibinfo {volume} {20}},\ \bibinfo {pages} {151} (\bibinfo {year}
  {1994})}\BibitemShut {NoStop}%
\bibitem [{\citenamefont {Holmes}\ \emph {et~al.}(2005)\citenamefont {Holmes},
  \citenamefont {Drinkwater},\ and\ \citenamefont {Wilcox}}]{Holmes2005}%
  \BibitemOpen
  \bibfield  {author} {\bibinfo {author} {\bibfnamefont {C.}~\bibnamefont
  {Holmes}}, \bibinfo {author} {\bibfnamefont {B.~W.}\ \bibnamefont
  {Drinkwater}},\ and\ \bibinfo {author} {\bibfnamefont {P.~D.}\ \bibnamefont
  {Wilcox}},\ }\bibfield  {title} {\bibinfo {title} {{Post-processing of the full matrix of ultrasonic transmit--receive array data for non-destructive evaluation}},\ }\href@noop {} {\bibfield  {journal} {\bibinfo  {journal} {NDT
  \& E Int.}\ }\textbf {\bibinfo {volume} {38}},\ \bibinfo {pages} {701}
  (\bibinfo {year} {2005})}\BibitemShut {NoStop}%
\bibitem [{\citenamefont {Watanabe}(2014)}]{Watanabe}%
  \BibitemOpen
  \bibfield  {author} {\bibinfo {author} {\bibfnamefont {K.}~\bibnamefont
  {Watanabe}},\ }\href@noop {} {\emph {\bibinfo {title} {Integral transform
  techniques for Green's functions. Chapter 2: Green's Functions for Laplace
  and Wave Equations}}}\ (\bibinfo  {publisher} {Springer, Cham, Switzerland},\
  \bibinfo {year} {2014})\BibitemShut {NoStop}%
\bibitem [{\citenamefont {Goodman}(1996)}]{Goodman1996}%
  \BibitemOpen
  \bibfield  {author} {\bibinfo {author} {\bibfnamefont {J.~W.}\ \bibnamefont
  {Goodman}},\ }\href@noop {} {\emph {\bibinfo {title} {{Introduction to
  Fourier Optics}}}},\ edited by\ \bibinfo {editor} {\bibfnamefont {S.~W.}\
  \bibnamefont {Director}}\ (\bibinfo  {publisher} {McGraw-Hill, Inc.},\
  \bibinfo {year} {1996})\ p.\ \bibinfo {pages} {491}\BibitemShut {NoStop}%
\bibitem [{\citenamefont {Born}\ and\ \citenamefont {Wolf}(2003)}]{Born}%
  \BibitemOpen
  \bibfield  {author} {\bibinfo {author} {\bibfnamefont {M.}~\bibnamefont
  {Born}}\ and\ \bibinfo {author} {\bibfnamefont {E.}~\bibnamefont {Wolf}},\
  }\href@noop {} {\emph {\bibinfo {title} {Principles of optics (Seventh
  edition)}}}\ (\bibinfo  {publisher} {Cambridge University Press, Cambridge},\
  \bibinfo {year} {2003})\BibitemShut {NoStop}%
\bibitem [{\citenamefont {Mehta}\ \emph {et~al.}(2008)\citenamefont {Mehta},
  \citenamefont {Thomas}, \citenamefont {BEll}, \citenamefont {Johnston},\ and\
  \citenamefont {Taylor-Robinson}}]{Mehta2008}%
  \BibitemOpen
  \bibfield  {author} {\bibinfo {author} {\bibfnamefont {S.~R.}\ \bibnamefont
  {Mehta}}, \bibinfo {author} {\bibfnamefont {E.~L.}\ \bibnamefont {Thomas}},
  \bibinfo {author} {\bibfnamefont {J.~D.}\ \bibnamefont {BEll}}, \bibinfo
  {author} {\bibfnamefont {D.~G.}\ \bibnamefont {Johnston}},\ and\ \bibinfo
  {author} {\bibfnamefont {S.~D.}\ \bibnamefont {Taylor-Robinson}},\ }\bibfield
   {title} {\bibinfo {title} {{Non-invasive means of measuring hepatic fat
  content}},\ }\href@noop {} {\bibfield  {journal} {\bibinfo  {journal} {World
  J Gastroenterol}\ }\textbf {\bibinfo {volume} {14}},\ \bibinfo {pages} {3476
  } (\bibinfo {year} {2008})}\BibitemShut {NoStop}%
\bibitem [{\citenamefont {Dasarathy}\ \emph {et~al.}(2009)\citenamefont
  {Dasarathy}, \citenamefont {Dasarathy}, \citenamefont {Khiyami},
  \citenamefont {Joseph}, \citenamefont {Lopez},\ and\ \citenamefont
  {McCullough}}]{Dasarathy2009}%
  \BibitemOpen
  \bibfield  {author} {\bibinfo {author} {\bibfnamefont {S.}~\bibnamefont
  {Dasarathy}}, \bibinfo {author} {\bibfnamefont {J.}~\bibnamefont
  {Dasarathy}}, \bibinfo {author} {\bibfnamefont {A.}~\bibnamefont {Khiyami}},
  \bibinfo {author} {\bibfnamefont {R.}~\bibnamefont {Joseph}}, \bibinfo
  {author} {\bibfnamefont {R.}~\bibnamefont {Lopez}},\ and\ \bibinfo {author}
  {\bibfnamefont {A.~J.}\ \bibnamefont {McCullough}},\ }\bibfield  {title}
  {\bibinfo {title} {{Validity of real time ultrasound in the diagnosis of
  hepatic steatosis: A prospective study}},\ }\href@noop {} {\bibfield
  {journal} {\bibinfo  {journal} {Hepatology}\ }\textbf {\bibinfo {volume}
  {51}},\ \bibinfo {pages} {1061 } (\bibinfo {year} {2009})}\BibitemShut
  {NoStop}%
\bibitem [{\citenamefont {Zubajlo}\ \emph {et~al.}(2018)\citenamefont
  {Zubajlo}, \citenamefont {Benjamin}, \citenamefont {Grajo}, \citenamefont
  {Kaliannan}, \citenamefont {Kang}, \citenamefont {Bhan}, \citenamefont
  {Thomenius}, \citenamefont {Anthony}, \citenamefont {Dhyani},\ and\
  \citenamefont {Samir}}]{Zubajlo2018}%
  \BibitemOpen
  \bibfield  {author} {\bibinfo {author} {\bibfnamefont {R.~E.}\ \bibnamefont
  {Zubajlo}}, \bibinfo {author} {\bibfnamefont {A.}~\bibnamefont {Benjamin}},
  \bibinfo {author} {\bibfnamefont {J.~R.}\ \bibnamefont {Grajo}}, \bibinfo
  {author} {\bibfnamefont {K.}~\bibnamefont {Kaliannan}}, \bibinfo {author}
  {\bibfnamefont {J.~X.}\ \bibnamefont {Kang}}, \bibinfo {author}
  {\bibfnamefont {A.~K.}\ \bibnamefont {Bhan}}, \bibinfo {author}
  {\bibfnamefont {K.~E.}\ \bibnamefont {Thomenius}}, \bibinfo {author}
  {\bibfnamefont {B.~W.}\ \bibnamefont {Anthony}}, \bibinfo {author}
  {\bibfnamefont {M.}~\bibnamefont {Dhyani}},\ and\ \bibinfo {author}
  {\bibfnamefont {A.~E.}\ \bibnamefont {Samir}},\ }\bibfield  {title} {\bibinfo
  {title} {{Experimental Validation of Longitudinal Speed of Sound Estimates in
  the Diagnosis of Hepatic Steatosis (Part II)}},\ }\href@noop {} {\bibfield
  {journal} {\bibinfo  {journal} {Ultrasound Med Biol}\ }\textbf {\bibinfo
  {volume} {44}},\ \bibinfo {pages} {2749 } (\bibinfo {year}
  {2018})}\BibitemShut {NoStop}%
\bibitem [{\citenamefont {Kennedy}(2005)}]{Kennedy2005}%
  \BibitemOpen
  \bibfield  {author} {\bibinfo {author} {\bibfnamefont {J.~E.}\ \bibnamefont
  {Kennedy}},\ }\bibfield  {title} {\bibinfo {title} {High-intensity focused
  ultrasound in the treatment of solid tumours},\ }\href@noop {} {\bibfield
  {journal} {\bibinfo  {journal} {Nat. Rev. Cancer}\ }\textbf {\bibinfo
  {volume} {5}},\ \bibinfo {pages} {321} (\bibinfo {year} {2005})}\BibitemShut
  {NoStop}%
\bibitem [{\citenamefont {Blackmore}\ \emph {et~al.}(2019)\citenamefont
  {Blackmore}, \citenamefont {Shrivastava}, \citenamefont {Sallet},
  \citenamefont {Butler},\ and\ \citenamefont {Cleveland}}]{Blackmore2019}%
  \BibitemOpen
  \bibfield  {author} {\bibinfo {author} {\bibfnamefont {J.}~\bibnamefont
  {Blackmore}}, \bibinfo {author} {\bibfnamefont {S.}~\bibnamefont
  {Shrivastava}}, \bibinfo {author} {\bibfnamefont {J.}~\bibnamefont {Sallet}},
  \bibinfo {author} {\bibfnamefont {C.}~\bibnamefont {Butler}},\ and\ \bibinfo
  {author} {\bibfnamefont {R.~O.}\ \bibnamefont {Cleveland}},\ }\bibfield
  {title} {\bibinfo {title} {Ultrasound neuromodulation: A review of results,
  mechanisms and safety},\ }\href@noop {} {\bibfield  {journal} {\bibinfo
  {journal} {Ultrasound Med. Biol.}\ }\textbf {\bibinfo {volume} {45}},\
  \bibinfo {pages} {1509} (\bibinfo {year} {2019})}\BibitemShut {NoStop}%
\bibitem [{\citenamefont {Macoskey}\ \emph {et~al.}(2018)\citenamefont
  {Macoskey}, \citenamefont {Hall}, \citenamefont {Sukovich}, \citenamefont
  {Choi}, \citenamefont {Ives}, \citenamefont {Johnsen}, \citenamefont {Cain},\
  and\ \citenamefont {Xu}}]{Macoskey2018}%
  \BibitemOpen
  \bibfield  {author} {\bibinfo {author} {\bibfnamefont {J.~J.}\ \bibnamefont
  {Macoskey}}, \bibinfo {author} {\bibfnamefont {T.~L.}\ \bibnamefont {Hall}},
  \bibinfo {author} {\bibfnamefont {J.~R.}\ \bibnamefont {Sukovich}}, \bibinfo
  {author} {\bibfnamefont {S.~W.}\ \bibnamefont {Choi}}, \bibinfo {author}
  {\bibfnamefont {K.}~\bibnamefont {Ives}}, \bibinfo {author} {\bibfnamefont
  {E.}~\bibnamefont {Johnsen}}, \bibinfo {author} {\bibfnamefont {C.~A.}\
  \bibnamefont {Cain}},\ and\ \bibinfo {author} {\bibfnamefont
  {Z.}~\bibnamefont {Xu}},\ }\bibfield  {title} {\bibinfo {title} {Soft-tissue
  aberration correction for histotripsy},\ }\href@noop {} {\bibfield  {journal}
  {\bibinfo  {journal} {IEEE Trans. Ultrason. Ferroelectr. Freq. Control.}\
  }\textbf {\bibinfo {volume} {65}},\ \bibinfo {pages} {2073 } (\bibinfo {year}
  {2018})}\BibitemShut {NoStop}%
\bibitem [{\citenamefont {Montaldo}\ \emph {et~al.}(2011)\citenamefont
  {Montaldo}, \citenamefont {Tanter},\ and\ \citenamefont
  {Fink}}]{Montaldo2011}%
  \BibitemOpen
  \bibfield  {author} {\bibinfo {author} {\bibfnamefont {G.}~\bibnamefont
  {Montaldo}}, \bibinfo {author} {\bibfnamefont {M.}~\bibnamefont {Tanter}},\
  and\ \bibinfo {author} {\bibfnamefont {M.}~\bibnamefont {Fink}},\ }\bibfield
  {title} {\bibinfo {title} {Time reversal of speckle noise},\ }\href@noop {}
  {\bibfield  {journal} {\bibinfo  {journal} {Phys. Rev. Lett.}\ }\textbf
  {\bibinfo {volume} {106}},\ \bibinfo {pages} {054301} (\bibinfo {year}
  {2011})}\BibitemShut {NoStop}%
\bibitem [{\citenamefont {Dahl}\ \emph {et~al.}(2005)\citenamefont {Dahl},
  \citenamefont {Soo},\ and\ \citenamefont {Trahey}}]{Dahl2005}%
  \BibitemOpen
  \bibfield  {author} {\bibinfo {author} {\bibfnamefont {J.~J.}\ \bibnamefont
  {Dahl}}, \bibinfo {author} {\bibfnamefont {M.~S.}\ \bibnamefont {Soo}},\ and\
  \bibinfo {author} {\bibfnamefont {G.~E.}\ \bibnamefont {Trahey}},\ }\bibfield
   {title} {\bibinfo {title} {Spatial and temporal aberrator stability for
  real-time adaptive imaging.},\ }\href@noop {} {\bibfield  {journal} {\bibinfo
   {journal} {IEEE Trans. Ultrason. Ferroelectr. Freq. Control}\ }\textbf
  {\bibinfo {volume} {52}},\ \bibinfo {pages} {1504} (\bibinfo {year}
  {2005})}\BibitemShut {NoStop}%
\bibitem [{\citenamefont {Kuga}\ and\ \citenamefont
  {Ishimaru}(1984)}]{Kuga1984}%
  \BibitemOpen
  \bibfield  {author} {\bibinfo {author} {\bibfnamefont {Y.}~\bibnamefont
  {Kuga}}\ and\ \bibinfo {author} {\bibfnamefont {A.}~\bibnamefont
  {Ishimaru}},\ }\bibfield  {title} {\bibinfo {title} {Retroreflectance from a
  dense distribution of spherical particles},\ }\href@noop {} {\bibfield
  {journal} {\bibinfo  {journal} {J. Opt. Soc. Am. A}\ }\textbf {\bibinfo
  {volume} {1}},\ \bibinfo {pages} {831} (\bibinfo {year} {1984})}\BibitemShut
  {NoStop}%
\bibitem [{\citenamefont {Van~Albada}\ and\ \citenamefont
  {Lagendijk}(1985)}]{VanAlbada1985b}%
  \BibitemOpen
  \bibfield  {author} {\bibinfo {author} {\bibfnamefont {M.~P.}\ \bibnamefont
  {Van~Albada}}\ and\ \bibinfo {author} {\bibfnamefont {A.}~\bibnamefont
  {Lagendijk}},\ }\bibfield  {title} {\bibinfo {title} {Observation of weak
  localization of light in a random medium},\ }\href@noop {} {\bibfield
  {journal} {\bibinfo  {journal} {Phys. Rev. Lett.}\ }\textbf {\bibinfo
  {volume} {55}},\ \bibinfo {pages} {2692} (\bibinfo {year}
  {1985})}\BibitemShut {NoStop}%
\bibitem [{\citenamefont {Wolf}\ and\ \citenamefont {Maret}(1985)}]{Wolf1985}%
  \BibitemOpen
  \bibfield  {author} {\bibinfo {author} {\bibfnamefont {P.~E.}\ \bibnamefont
  {Wolf}}\ and\ \bibinfo {author} {\bibfnamefont {G.}~\bibnamefont {Maret}},\
  }\bibfield  {title} {\bibinfo {title} {Weak localization and coherent
  backscattering of photons in disordered media},\ }\href@noop {} {\bibfield
  {journal} {\bibinfo  {journal} {Phys. Rev. Lett.}\ }\textbf {\bibinfo
  {volume} {55}},\ \bibinfo {pages} {2696} (\bibinfo {year}
  {1985})}\BibitemShut {NoStop}%
\bibitem [{\citenamefont {Akkermans}\ \emph {et~al.}(1988)\citenamefont
  {Akkermans}, \citenamefont {Wolf}, \citenamefont {Maynard},\ and\
  \citenamefont {Maret}}]{Akkermans1988c}%
  \BibitemOpen
  \bibfield  {author} {\bibinfo {author} {\bibfnamefont {E.}~\bibnamefont
  {Akkermans}}, \bibinfo {author} {\bibfnamefont {P.~E.}\ \bibnamefont {Wolf}},
  \bibinfo {author} {\bibfnamefont {R.}~\bibnamefont {Maynard}},\ and\ \bibinfo
  {author} {\bibfnamefont {G.}~\bibnamefont {Maret}},\ }\bibfield  {title}
  {\bibinfo {title} {{Theoretical study of the coherent backscattering of light
  by disordered media}},\ }\href@noop {} {\bibfield  {journal} {\bibinfo
  {journal} {J. Phys. France}\ }\textbf {\bibinfo {volume} {49}},\ \bibinfo
  {pages} {77} (\bibinfo {year} {1988})}\BibitemShut {NoStop}%
\bibitem [{\citenamefont {Bayer}\ and\ \citenamefont
  {Niederdr{\"{a}}nk}(1993)}]{Bayer1993a}%
  \BibitemOpen
  \bibfield  {author} {\bibinfo {author} {\bibfnamefont {G.}~\bibnamefont
  {Bayer}}\ and\ \bibinfo {author} {\bibfnamefont {T.}~\bibnamefont
  {Niederdr{\"{a}}nk}},\ }\bibfield  {title} {\bibinfo {title} {Weak
  localization of acoustic waves in strongly scattering media},\ }\href@noop {}
  {\bibfield  {journal} {\bibinfo  {journal} {Phys. Rev. Lett.}\ }\textbf
  {\bibinfo {volume} {70}},\ \bibinfo {pages} {3884} (\bibinfo {year}
  {1993})}\BibitemShut {NoStop}%
\bibitem [{\citenamefont {Tourin}\ \emph {et~al.}(1997)\citenamefont {Tourin},
  \citenamefont {Derode}, \citenamefont {Roux}, \citenamefont {van Tiggelen},\
  and\ \citenamefont {Fink}}]{Tourin1997}%
  \BibitemOpen
  \bibfield  {author} {\bibinfo {author} {\bibfnamefont {A.}~\bibnamefont
  {Tourin}}, \bibinfo {author} {\bibfnamefont {A.}~\bibnamefont {Derode}},
  \bibinfo {author} {\bibfnamefont {P.}~\bibnamefont {Roux}}, \bibinfo {author}
  {\bibfnamefont {B.~A.}\ \bibnamefont {van Tiggelen}},\ and\ \bibinfo {author}
  {\bibfnamefont {M.}~\bibnamefont {Fink}},\ }\bibfield  {title} {\bibinfo
  {title} {Time-dependent coherent backscattering of acoustic waves},\
  }\href@noop {} {\bibfield  {journal} {\bibinfo  {journal} {Phys. Rev. Lett.}\
  }\textbf {\bibinfo {volume} {79}},\ \bibinfo {pages} {3637} (\bibinfo {year}
  {1997})}\BibitemShut {NoStop}%
\bibitem [{\citenamefont {Larose}\ \emph {et~al.}(2004)\citenamefont {Larose},
  \citenamefont {Margerin}, \citenamefont {van Tiggelen},\ and\ \citenamefont
  {Campillo}}]{Larose2004}%
  \BibitemOpen
  \bibfield  {author} {\bibinfo {author} {\bibfnamefont {E.}~\bibnamefont
  {Larose}}, \bibinfo {author} {\bibfnamefont {L.}~\bibnamefont {Margerin}},
  \bibinfo {author} {\bibfnamefont {B.~A.}\ \bibnamefont {van Tiggelen}},\ and\
  \bibinfo {author} {\bibfnamefont {M.}~\bibnamefont {Campillo}},\ }\bibfield
  {title} {\bibinfo {title} {Weak localization of seismic waves},\ }\href@noop
  {} {\bibfield  {journal} {\bibinfo  {journal} {Phys. Rev. Lett.}\ }\textbf
  {\bibinfo {volume} {93}},\ \bibinfo {pages} {048501} (\bibinfo {year}
  {2004})}\BibitemShut {NoStop}%
\bibitem [{\citenamefont {Aubry}\ and\ \citenamefont
  {Derode}(2007)}]{Aubry2007b}%
  \BibitemOpen
  \bibfield  {author} {\bibinfo {author} {\bibfnamefont {A.}~\bibnamefont
  {Aubry}}\ and\ \bibinfo {author} {\bibfnamefont {A.}~\bibnamefont {Derode}},\
  }\bibfield  {title} {\bibinfo {title} {Ultrasonic imaging of highly
  scattering media from local measurements of the diffusion constant:
  Separation of coherent and incoherent intensities},\ }\href@noop {}
  {\bibfield  {journal} {\bibinfo  {journal} {Phys. Rev. E}\ }\textbf {\bibinfo
  {volume} {75}},\ \bibinfo {pages} {026602} (\bibinfo {year}
  {2007})}\BibitemShut {NoStop}%
\bibitem [{\citenamefont {Sheng}(2006)}]{Sheng2006}%
  \BibitemOpen
  \bibfield  {author} {\bibinfo {author} {\bibfnamefont {P.}~\bibnamefont
  {Sheng}},\ }\href@noop {} {\emph {\bibinfo {title} {{Introduction to Wave
  Scattering, Localization and Mesoscopic Phenomena}}}},\ \bibinfo {edition}
  {second edi}\ ed.,\ edited by\ \bibinfo {editor} {\bibfnamefont
  {R.}~\bibnamefont {Hull}}, \bibinfo {editor} {\bibfnamefont {R.~M.~J.}\
  \bibnamefont {Osgood}}, \bibinfo {editor} {\bibfnamefont {J.}~\bibnamefont
  {Parisi}},\ and\ \bibinfo {editor} {\bibfnamefont {H.}~\bibnamefont
  {Warlimont}},\ Vol.\ \bibinfo {volume} {468}\ (\bibinfo  {publisher}
  {Springer},\ \bibinfo {address} {Berlin},\ \bibinfo {year} {2006})\ p.\
  \bibinfo {pages} {333}\BibitemShut {NoStop}%
\bibitem [{\citenamefont {Velichko}(2019)}]{Velichko2019}%
  \BibitemOpen
  \bibfield  {author} {\bibinfo {author} {\bibfnamefont {A.}~\bibnamefont
  {Velichko}},\ }\bibfield  {title} {\bibinfo {title} {Quantification of the
  effect of multiple scattering on array imaging performance},\ }\href@noop {}
  {\bibfield  {journal} {\bibinfo  {journal} {IEEE Trans. Ultrason.
  Ferroelectr. Freq. Control.}\ }\textbf {\bibinfo {volume} {67}},\ \bibinfo
  {pages} {1} (\bibinfo {year} {2019})}\BibitemShut {NoStop}%
\bibitem [{\citenamefont {Jonckheere}\ \emph {et~al.}(2000)\citenamefont
  {Jonckheere}, \citenamefont {M{\"{u}}ller}, \citenamefont {Kaiser},
  \citenamefont {Miniatura},\ and\ \citenamefont {Delande}}]{Jonckheere2000}%
  \BibitemOpen
  \bibfield  {author} {\bibinfo {author} {\bibfnamefont {T.}~\bibnamefont
  {Jonckheere}}, \bibinfo {author} {\bibfnamefont {C.~A.}\ \bibnamefont
  {M{\"{u}}ller}}, \bibinfo {author} {\bibfnamefont {R.}~\bibnamefont
  {Kaiser}}, \bibinfo {author} {\bibfnamefont {C.}~\bibnamefont {Miniatura}},\
  and\ \bibinfo {author} {\bibfnamefont {D.}~\bibnamefont {Delande}},\
  }\bibfield  {title} {\bibinfo {title} {{Multiple scattering of light by atoms
  in the weak localization regime}},\ }\href@noop {} {\bibfield  {journal}
  {\bibinfo  {journal} {Phys. Rev. Lett.}\ }\textbf {\bibinfo {volume} {85}},\
  \bibinfo {pages} {4269} (\bibinfo {year} {2000})}\BibitemShut {NoStop}%
\bibitem [{\citenamefont {Wolf}\ \emph {et~al.}(1988)\citenamefont {Wolf},
  \citenamefont {Maret}, \citenamefont {Akkermans},\ and\ \citenamefont
  {Maynard}}]{Wolf1988c}%
  \BibitemOpen
  \bibfield  {author} {\bibinfo {author} {\bibfnamefont {P.~E.}\ \bibnamefont
  {Wolf}}, \bibinfo {author} {\bibfnamefont {G.}~\bibnamefont {Maret}},
  \bibinfo {author} {\bibfnamefont {E.}~\bibnamefont {Akkermans}},\ and\
  \bibinfo {author} {\bibfnamefont {R.}~\bibnamefont {Maynard}},\ }\bibfield
  {title} {\bibinfo {title} {{Optical coherent backscattering by random media:
  an experimental study}},\ }\href@noop {} {\bibfield  {journal} {\bibinfo
  {journal} {J. Phys. France}\ }\textbf {\bibinfo {volume} {49}},\ \bibinfo
  {pages} {63} (\bibinfo {year} {1988})}\BibitemShut {NoStop}%
\bibitem [{\citenamefont {Cobus}\ \emph {et~al.}(2017)\citenamefont {Cobus},
  \citenamefont {van Tiggelen}, \citenamefont {Derode},\ and\ \citenamefont
  {Page}}]{Cobus2017}%
  \BibitemOpen
  \bibfield  {author} {\bibinfo {author} {\bibfnamefont {L.~A.}\ \bibnamefont
  {Cobus}}, \bibinfo {author} {\bibfnamefont {B.~A.}\ \bibnamefont {van
  Tiggelen}}, \bibinfo {author} {\bibfnamefont {A.}~\bibnamefont {Derode}},\
  and\ \bibinfo {author} {\bibfnamefont {J.~H.}\ \bibnamefont {Page}},\
  }\bibfield  {title} {\bibinfo {title} {{Dynamic coherent backscattering of
  ultrasound in three-dimensional strongly-scattering media}},\ }\href
  {http://link.springer.com/10.1140/epjst/e2016-60340-3} {\bibfield  {journal}
  {\bibinfo  {journal} {Eur. Phys. J. ST}\ }\textbf {\bibinfo {volume} {226}},\
  \bibinfo {pages} {1549} (\bibinfo {year} {2017})}\BibitemShut {NoStop}%
\bibitem [{\citenamefont {Badon}\ \emph
  {et~al.}(2016{\natexlab{b}})\citenamefont {Badon}, \citenamefont {Li},
  \citenamefont {Lerosey}, \citenamefont {Boccara}, \citenamefont {Fink},\ and\
  \citenamefont {Aubry}}]{Badon2016b}%
  \BibitemOpen
  \bibfield  {author} {\bibinfo {author} {\bibfnamefont {A.}~\bibnamefont
  {Badon}}, \bibinfo {author} {\bibfnamefont {D.}~\bibnamefont {Li}}, \bibinfo
  {author} {\bibfnamefont {G.}~\bibnamefont {Lerosey}}, \bibinfo {author}
  {\bibfnamefont {A.~C.}\ \bibnamefont {Boccara}}, \bibinfo {author}
  {\bibfnamefont {M.}~\bibnamefont {Fink}},\ and\ \bibinfo {author}
  {\bibfnamefont {A.}~\bibnamefont {Aubry}},\ }\bibfield  {title} {\bibinfo
  {title} {Spatio-temporal imaging of light transport in highly scattering
  media under white light illumination},\ }\href@noop {} {\bibfield  {journal}
  {\bibinfo  {journal} {Optica}\ }\textbf {\bibinfo {volume} {3}},\ \bibinfo
  {pages} {1160} (\bibinfo {year} {2016}{\natexlab{b}})}\BibitemShut {NoStop}%
\bibitem [{\citenamefont {Moran}\ \emph {et~al.}(1995)\citenamefont {Moran},
  \citenamefont {Bush},\ and\ \citenamefont {Bamber}}]{Moran1995}%
  \BibitemOpen
  \bibfield  {author} {\bibinfo {author} {\bibfnamefont {C.~M.}\ \bibnamefont
  {Moran}}, \bibinfo {author} {\bibfnamefont {N.~L.}\ \bibnamefont {Bush}},\
  and\ \bibinfo {author} {\bibfnamefont {J.~C.}\ \bibnamefont {Bamber}},\
  }\bibfield  {title} {\bibinfo {title} {Ultrasonic propagation properties of
  excised human skin},\ }\href@noop {} {\bibfield  {journal} {\bibinfo
  {journal} {Ultrasound Med. Biol.}\ }\textbf {\bibinfo {volume} {71}},\
  \bibinfo {pages} {1177} (\bibinfo {year} {1995})}\BibitemShut {NoStop}%
\bibitem [{\citenamefont {Errabolu}\ \emph {et~al.}(1988)\citenamefont
  {Errabolu}, \citenamefont {Sehgala}, \citenamefont {Bahn},\ and\
  \citenamefont {Greenleaf}}]{Errabolu1998}%
  \BibitemOpen
  \bibfield  {author} {\bibinfo {author} {\bibfnamefont {R.~L.}\ \bibnamefont
  {Errabolu}}, \bibinfo {author} {\bibfnamefont {C.~M.}\ \bibnamefont
  {Sehgala}}, \bibinfo {author} {\bibfnamefont {R.~C.}\ \bibnamefont {Bahn}},\
  and\ \bibinfo {author} {\bibfnamefont {J.~F.}\ \bibnamefont {Greenleaf}},\
  }\bibfield  {title} {\bibinfo {title} {Measurement of ultrasonic nonlinear
  parameter in excised fat tissues},\ }\href@noop {} {\bibfield  {journal}
  {\bibinfo  {journal} {Ultrasound Med. Biol.}\ }\textbf {\bibinfo {volume}
  {14}},\ \bibinfo {pages} {137 } (\bibinfo {year} {1988})}\BibitemShut
  {NoStop}%
\bibitem [{\citenamefont {Rajagopalan}\ \emph {et~al.}(1979)\citenamefont
  {Rajagopalan}, \citenamefont {Greenleaf}, \citenamefont {Thomas},
  \citenamefont {Johnson},\ and\ \citenamefont {Bahn}}]{Rajagopalan1979}%
  \BibitemOpen
  \bibfield  {author} {\bibinfo {author} {\bibfnamefont {B.}~\bibnamefont
  {Rajagopalan}}, \bibinfo {author} {\bibfnamefont {J.~F.}\ \bibnamefont
  {Greenleaf}}, \bibinfo {author} {\bibfnamefont {P.~J.}\ \bibnamefont
  {Thomas}}, \bibinfo {author} {\bibfnamefont {S.~A.}\ \bibnamefont
  {Johnson}},\ and\ \bibinfo {author} {\bibfnamefont {R.~C.}\ \bibnamefont
  {Bahn}},\ }\bibfield  {title} {\bibinfo {title} {Variation of acoustic speed
  with temperature in various exised human tissues studied by ultrasound
  computerized tomography},\ }in\ \href@noop {} {\emph {\bibinfo {booktitle}
  {Ultrasonic Tissue Characterization II}}},\ \bibinfo {editor} {edited by\
  \bibinfo {editor} {\bibnamefont {Linzer}}}\ (\bibinfo  {publisher} {U.S.
  Department of Commerce},\ \bibinfo {year} {1979})\ pp.\ \bibinfo {pages} {227
  -- 233}\BibitemShut {NoStop}%
\bibitem [{\citenamefont {Lin}\ \emph {et~al.}(1987)\citenamefont {Lin},
  \citenamefont {Ophir},\ and\ \citenamefont {Potter}}]{Lin1987}%
  \BibitemOpen
  \bibfield  {author} {\bibinfo {author} {\bibfnamefont {T.}~\bibnamefont
  {Lin}}, \bibinfo {author} {\bibfnamefont {J.}~\bibnamefont {Ophir}},\ and\
  \bibinfo {author} {\bibfnamefont {G.}~\bibnamefont {Potter}},\ }\bibfield
  {title} {\bibinfo {title} {Correlations of sound speed with tissue
  constituents in normal and diffuse liver disease},\ }\href@noop {} {\bibfield
   {journal} {\bibinfo  {journal} {Ultrason. Imaging}\ }\textbf {\bibinfo
  {volume} {9}},\ \bibinfo {pages} {29 } (\bibinfo {year} {1987})}\BibitemShut
  {NoStop}%
\bibitem [{\citenamefont {Boozari}\ \emph {et~al.}(2010)\citenamefont
  {Boozari}, \citenamefont {Botthoff}, \citenamefont {Mederacke}, \citenamefont
  {Hahn}, \citenamefont {Reising}, \citenamefont {Rifai}, \citenamefont
  {Wedemeyer}, \citenamefont {Bahr}, \citenamefont {Kubicka}, \citenamefont
  {Manns},\ and\ \citenamefont {Gebel}}]{Boozari2010}%
  \BibitemOpen
  \bibfield  {author} {\bibinfo {author} {\bibfnamefont {B.}~\bibnamefont
  {Boozari}}, \bibinfo {author} {\bibfnamefont {A.}~\bibnamefont {Botthoff}},
  \bibinfo {author} {\bibfnamefont {I.}~\bibnamefont {Mederacke}}, \bibinfo
  {author} {\bibfnamefont {A.}~\bibnamefont {Hahn}}, \bibinfo {author}
  {\bibfnamefont {A.}~\bibnamefont {Reising}}, \bibinfo {author} {\bibfnamefont
  {K.}~\bibnamefont {Rifai}}, \bibinfo {author} {\bibfnamefont
  {H.}~\bibnamefont {Wedemeyer}}, \bibinfo {author} {\bibfnamefont
  {M.}~\bibnamefont {Bahr}}, \bibinfo {author} {\bibfnamefont {s.}~\bibnamefont
  {Kubicka}}, \bibinfo {author} {\bibfnamefont {M.}~\bibnamefont {Manns}},\
  and\ \bibinfo {author} {\bibfnamefont {M.}~\bibnamefont {Gebel}},\ }\bibfield
   {title} {\bibinfo {title} {Evaluation of sound speed for detection of liver
  fibrosis: prospective comparison with transient dynamic elastography and
  histology},\ }\href@noop {} {\bibfield  {journal} {\bibinfo  {journal} {J.
  Ultrasound Med.}\ }\textbf {\bibinfo {volume} {29}},\ \bibinfo {pages} {1581
  } (\bibinfo {year} {2010})}\BibitemShut {NoStop}%
\bibitem [{\citenamefont {Jakovljevic}\ \emph {et~al.}(2018)\citenamefont
  {Jakovljevic}, \citenamefont {Hsieh}, \citenamefont {Ali}, \citenamefont
  {Kung}, \citenamefont {Hyun},\ and\ \citenamefont {Dahl}}]{Jakovljevic2018}%
  \BibitemOpen
  \bibfield  {author} {\bibinfo {author} {\bibfnamefont {M.}~\bibnamefont
  {Jakovljevic}}, \bibinfo {author} {\bibfnamefont {S.}~\bibnamefont {Hsieh}},
  \bibinfo {author} {\bibfnamefont {R.}~\bibnamefont {Ali}}, \bibinfo {author}
  {\bibfnamefont {G.~C. L.~K.}\ \bibnamefont {Kung}}, \bibinfo {author}
  {\bibfnamefont {D.}~\bibnamefont {Hyun}},\ and\ \bibinfo {author}
  {\bibfnamefont {J.~J.}\ \bibnamefont {Dahl}},\ }\bibfield  {title} {\bibinfo
  {title} {{Local speed of sound estimation in tissue using pulse-echo
  ultrasound: Model-based approach}},\ }\href@noop {} {\bibfield  {journal}
  {\bibinfo  {journal} {J. Acoust. Soc. Am.}\ }\textbf {\bibinfo {volume}
  {144}},\ \bibinfo {pages} {254} (\bibinfo {year} {2018})}\BibitemShut
  {NoStop}%
\bibitem [{\citenamefont {Franceschini}\ \emph {et~al.}(2019)\citenamefont
  {Franceschini}, \citenamefont {Escoffre}, \citenamefont {Novel},
  \citenamefont {Auboire}, \citenamefont {Mendes}, \citenamefont {Benane},
  \citenamefont {Bouakaz},\ and\ \citenamefont {Bassetz}}]{Francechini2019}%
  \BibitemOpen
  \bibfield  {author} {\bibinfo {author} {\bibfnamefont {E.}~\bibnamefont
  {Franceschini}}, \bibinfo {author} {\bibfnamefont {J.-M.}\ \bibnamefont
  {Escoffre}}, \bibinfo {author} {\bibfnamefont {A.}~\bibnamefont {Novel}},
  \bibinfo {author} {\bibfnamefont {L.}~\bibnamefont {Auboire}}, \bibinfo
  {author} {\bibfnamefont {V.}~\bibnamefont {Mendes}}, \bibinfo {author}
  {\bibfnamefont {Y.~M.}\ \bibnamefont {Benane}}, \bibinfo {author}
  {\bibfnamefont {A.}~\bibnamefont {Bouakaz}},\ and\ \bibinfo {author}
  {\bibfnamefont {O.}~\bibnamefont {Bassetz}},\ }\bibfield  {title} {\bibinfo
  {title} {Quantitative ultrasound in ex vivo fibrotic rabbit livers},\
  }\href@noop {} {\bibfield  {journal} {\bibinfo  {journal} {Ultrasound Med.
  Biol.}\ }\textbf {\bibinfo {volume} {45}},\ \bibinfo {pages} {1777 }
  (\bibinfo {year} {2019})}\BibitemShut {NoStop}%
\bibitem [{\citenamefont {Oelze}\ and\ \citenamefont
  {Mamou}(2016)}]{Oelze2016}%
  \BibitemOpen
  \bibfield  {author} {\bibinfo {author} {\bibfnamefont {M.~L.}\ \bibnamefont
  {Oelze}}\ and\ \bibinfo {author} {\bibfnamefont {J.}~\bibnamefont {Mamou}},\
  }\bibfield  {title} {\bibinfo {title} {Review of quantitative ultrasound:
  Envelope statistics and backscatter coefficient imaging and contributions to
  diagnostic ultrasound},\ }\href@noop {} {\bibfield  {journal} {\bibinfo
  {journal} {IEEE Trans. Ultrason. Ferroelectr. Freq. Control.}\ }\textbf
  {\bibinfo {volume} {63}} (\bibinfo {year} {2016})}\BibitemShut {NoStop}%
\bibitem [{\citenamefont {Mohanty}\ \emph {et~al.}(2018)\citenamefont
  {Mohanty}, \citenamefont {Blackwell}, \citenamefont {Masuodi}, \citenamefont
  {Ali}, \citenamefont {Egan},\ and\ \citenamefont {Muller}}]{Mohanty2018}%
  \BibitemOpen
  \bibfield  {author} {\bibinfo {author} {\bibfnamefont {K.}~\bibnamefont
  {Mohanty}}, \bibinfo {author} {\bibfnamefont {J.}~\bibnamefont {Blackwell}},
  \bibinfo {author} {\bibfnamefont {S.~B.}\ \bibnamefont {Masuodi}}, \bibinfo
  {author} {\bibfnamefont {M.~H.}\ \bibnamefont {Ali}}, \bibinfo {author}
  {\bibfnamefont {T.}~\bibnamefont {Egan}},\ and\ \bibinfo {author}
  {\bibfnamefont {M.}~\bibnamefont {Muller}},\ }\bibfield  {title} {\bibinfo
  {title} {{1-Dimensional} quantitative micro-architecture mapping of multiple
  scattering media using backscattering of ultrasound in the near-field:
  Application to nodule imaging in the lungs},\ }\href@noop {} {\bibfield
  {journal} {\bibinfo  {journal} {Appl. Phys. Lett}\ }\textbf {\bibinfo
  {volume} {113}} (\bibinfo {year} {2018})}\BibitemShut {NoStop}%
\bibitem [{\citenamefont {Long}\ \emph {et~al.}(2018)\citenamefont {Long},
  \citenamefont {Bottenus},\ and\ \citenamefont {Trahey}}]{Long2018}%
  \BibitemOpen
  \bibfield  {author} {\bibinfo {author} {\bibfnamefont {W.}~\bibnamefont
  {Long}}, \bibinfo {author} {\bibfnamefont {N.}~\bibnamefont {Bottenus}},\
  and\ \bibinfo {author} {\bibfnamefont {G.~E.}\ \bibnamefont {Trahey}},\
  }\bibfield  {title} {\bibinfo {title} {Lag-one coherence as a metric for
  ultrasonic image quality},\ }\href@noop {} {\bibfield  {journal} {\bibinfo
  {journal} {IEEE Trans. Ultrason. Ferroelectr. Freq. Control.}\ }\textbf
  {\bibinfo {volume} {65}},\ \bibinfo {pages} {1768 } (\bibinfo {year}
  {2018})}\BibitemShut {NoStop}%
\bibitem [{\citenamefont {Whittingham}(1999)}]{Whittingham}%
  \BibitemOpen
  \bibfield  {author} {\bibinfo {author} {\bibfnamefont {T.~A.}\ \bibnamefont
  {Whittingham}},\ }\bibfield  {title} {\bibinfo {title} {Tissue harmonic
  imaging},\ }\href@noop {} {\bibfield  {journal} {\bibinfo  {journal} {Eur.
  Radiol.}\ }\textbf {\bibinfo {volume} {9}},\ \bibinfo {pages} {S323}
  (\bibinfo {year} {1999})}\BibitemShut {NoStop}%
\bibitem [{\citenamefont {Tiran}\ \emph {et~al.}(2015)\citenamefont {Tiran},
  \citenamefont {Deffieux}, \citenamefont {Correia}, \citenamefont {Maresca},
  \citenamefont {Osmanski}, \citenamefont {Sieu}, \citenamefont {Bergel},
  \citenamefont {Cohen}, \citenamefont {Pernot},\ and\ \citenamefont
  {Tanter}}]{Tiran_2015}%
  \BibitemOpen
  \bibfield  {author} {\bibinfo {author} {\bibfnamefont {E.}~\bibnamefont
  {Tiran}}, \bibinfo {author} {\bibfnamefont {T.}~\bibnamefont {Deffieux}},
  \bibinfo {author} {\bibfnamefont {M.}~\bibnamefont {Correia}}, \bibinfo
  {author} {\bibfnamefont {D.}~\bibnamefont {Maresca}}, \bibinfo {author}
  {\bibfnamefont {B.-F.}\ \bibnamefont {Osmanski}}, \bibinfo {author}
  {\bibfnamefont {L.-A.}\ \bibnamefont {Sieu}}, \bibinfo {author}
  {\bibfnamefont {A.}~\bibnamefont {Bergel}}, \bibinfo {author} {\bibfnamefont
  {I.}~\bibnamefont {Cohen}}, \bibinfo {author} {\bibfnamefont
  {M.}~\bibnamefont {Pernot}},\ and\ \bibinfo {author} {\bibfnamefont
  {M.}~\bibnamefont {Tanter}},\ }\bibfield  {title} {\bibinfo {title}
  {Multiplane wave imaging increases signal-to-noise ratio in ultrafast
  ultrasound imaging},\ }\href {https://doi.org/10.1088/0031-9155/60/21/8549}
  {\bibfield  {journal} {\bibinfo  {journal} {Phys. Med. Biol.}\ }\textbf
  {\bibinfo {volume} {60}},\ \bibinfo {pages} {8549} (\bibinfo {year}
  {2015})}\BibitemShut {NoStop}%
\bibitem [{\citenamefont {Mahajan}(1982)}]{mahajan1982strehl}%
  \BibitemOpen
  \bibfield  {author} {\bibinfo {author} {\bibfnamefont {V.~N.}\ \bibnamefont
  {Mahajan}},\ }\bibfield  {title} {\bibinfo {title} {Strehl ratio for primary
  aberrations: some analytical results for circular and annular pupils},\
  }\href@noop {} {\bibfield  {journal} {\bibinfo  {journal} {J. Opt. Soc. Am.}\
  }\textbf {\bibinfo {volume} {72}},\ \bibinfo {pages} {1258} (\bibinfo {year}
  {1982})}\BibitemShut {NoStop}%
\end{thebibliography}

\end{document}